\documentclass[
	reprint,
	showkeys,
	superscriptaddress,
	aps,
	pra,
	longbibliography,
	nofootinbib
]{revtex4-2}

\usepackage[utf8]{inputenc}	

\usepackage{mathtools}	
\usepackage{amsmath}	
\usepackage{amssymb}	
\usepackage{bm}			
\usepackage{tensor}		
\usepackage{upgreek}	
\usepackage{dsfont}		
\usepackage{physics}	
\usepackage{slashed}	
\usepackage{cancel}		

\usepackage{comment}	
\usepackage{placeins}	

\usepackage{multirow}	
\usepackage{graphicx}	
\usepackage{dcolumn}	

\usepackage[hyperref]{xcolor}	
	\definecolor{goethe-blau}{cmyk}{1.0,0.2,0.0,0.4}
	\definecolor{hellgrau}{cmyk}{0.04,0.04,0.05,0.02}
	\definecolor{sandgrau}{cmyk}{0.12,0.09,0.13,0.0}
	\definecolor{dunkelgrau}{cmyk}{0.25,0.25,0.30,0.75}
	\definecolor{emo-rot}{cmyk}{0.04,1.0,0.8,0.07}
	\definecolor{purple}{cmyk}{0.08,1.0,0.3,0.36}
	\definecolor{senfgelb}{cmyk}{0.01,0.25,1.0,0.05}
	\definecolor{gruen}{cmyk}{0.62,0.4,0.87,0.09}
	\definecolor{magenta}{cmyk}{0.08,0.86,0.12,0.12}
	\definecolor{orange}{cmyk}{0.0,0.7,1.0,0.04}
	\definecolor{sonnengelb}{cmyk}{0.0,0.12,0.95,0.0}
	\definecolor{helles-gruen}{cmyk}{0.4,0.17,0.81,0.07}
	\definecolor{lichtblau}{cmyk}{0.8,0.0,0.06,0.04}

\usepackage[
	colorlinks,
	pdfpagelabels,
	breaklinks,
	pdfstartview=FitH,
	bookmarksopen=true,
	bookmarksnumbered=true,
	bookmarksopenlevel=2,
	plainpages=false,
	hypertexnames=false,
	citecolor=emo-rot,
	linkcolor=goethe-blau,
	urlcolor=purple,
	pdftitle={Revisiting the spatially inhomogeneous condensates in the (1+1)-dimensional chiral Gross-Neveu model via the bosonic two-point function in the infinite-N limit},
	pdfauthor={Marc Winstel, Adrian Koenigstein},
	pdfkeywords={chiral Gross-Neveu model, phase diagram, infinite-N limit, bosonic two-point function, stability analysis, inhomogeneous phase, quark chemical potential, chiral imbalance, temperature, chiral spiral},
]{hyperref}

\usepackage{bookmark}

\usepackage[capitalise,sort,compress]{cleveref}	

\allowdisplaybreaks				

\renewcommand{\S}{{\mathcal{S}}}					
\renewcommand{\dd}{{\mathrm{d}}}					
\newcommand{\barpsi}{{\bar{\psi}}}					
\newcommand{\ii}{{\mathrm{i}}}						
\newcommand{\ee}{{\mathrm{e}}}						
\newcommand{\Det}{\mathrm{Det}}						
\newcommand{\gammachiral}{{\gamma_{\mathrm{ch}}}}	
\newcommand{\gammaleft}{{\gamma_{\mathrm{L}}}}		
\newcommand{\gammaright}{{\gamma_{\mathrm{R}}}}		
\newcommand{\gammaleftright}{{\gamma_{\mathrm{L}/\mathrm{R}}}}		
\newcommand{\gammarightleft}{{\gamma_{\mathrm{R}/\mathrm{L}}}}		
\newcommand{\dimDirac}{{d_\gamma}}							
\newcommand{\ffcoupling}{\lambda}					
\newcommand{\ycoupling}{h}							
\newcommand{\muchiral}{{\tilde{\mu}}}				
\newcommand{\varphileft}{{\Delta^\ast}}	
\newcommand{\varphiright}{{\Delta}}	
\newcommand{\barvarphileft}{{\bar{\Delta}^\ast}}	
\newcommand{\barvarphiright}{{\bar{\Delta}}}	
\newcommand{\varphileftright}{{\Delta^{\ast/\phantom{\circ}}}}	
\newcommand{\varphirightleft}{{\Delta^{\phantom{\circ}/\ast}}}	
\newcommand{\nf}{{n_\mathrm{f}}}					

\newcommand{\vdistance}{\vphantom{\bigg(\bigg)}}	
\newcommand{\Vdistance}{\vphantom{\Bigg(\Bigg)}}	

\newcommand{\U}{\ensuremath{\mathrm{U}}}

\newcommand{\st}[1]{\ensuremath{\mathbf{#1}}}

\DeclareMathOperator{\sumint}{\, \underset{\st{p}}{\mathrlap{\sum}\int} \, \,}

\DeclareUnicodeCharacter{2A0B}{\sumint}

\usepackage[acronym]{glossaries-extra}	
	\glssetcategoryattribute{acronym}{nohyperfirst}{true}
	\setabbreviationstyle[acronym]{long-short}

	\newacronym[plural=QFTs,firstplural=quantum field theories]{qft}{QFT}{quantum field theory}
	\newacronym{qcd}{QCD}{Quantum Chromodynamics}
	\newacronym{frg}{FRG}{Functional Renormalization Group}
	\newacronym{njl}{NJL}{Nambu-Jona-Lasinio}
	\newacronym{gn}{GN}{Gross-Neveu}
	\newacronym{chgn}{$\chi$GN}{chiral Gross-Neveu}
	\newacronym{hbp}{HBP}{homogeneously broken phase}
	\newacronym{ip}{IP}{inhomogeneous phase}
	\newacronym{sp}{SP}{symmetric phase}
	\newacronym{cdw}{CDW}{chiral density wave}
	\newacronym{uv}{UV}{ultra violet}
	\newacronym{ir}{IR}{infra red}
	\newacronym{wrt}{w.r.t.}{with respect to}
	\newacronym{adpt}{ADPT}{almost degenerate perturbation theory}

\begin{document}



	\title{
		Revisiting the spatially inhomogeneous condensates in the $(1 + 1)$-dimensional chiral Gross-Neveu model via the bosonic two-point function in the infinite-$N$ limit
	}


	\author{Adrian Koenigstein}
		\email{adrian.koenigstein@uni-jena.de}
		\affiliation{
			Theoretisch-Physikalisches Institut,
			Friedrich-Schiller Universität,
			D-07743 Jena,
			Germany.
		}
	
		\author{Marc Winstel}
	\email{winstel@itp.uni-frankfurt.de, author to whom any correspondence should be addressed}
	\affiliation{
		Institut für Theoretische Physik,
		Johann Wolfgang Goethe-Universität,
		D-60438 Frankfurt am Main,
		Germany.
	}

	\date{\today}

	\begin{abstract}
		This work shows that the known phase boundary between the phase with chiral symmetry and the phase of spatially inhomogeneous chiral symmetry breaking in the phase diagram of the $(1 + 1)$-dimensional chiral Gross-Neveu model can be detected from the bosonic two-point function alone and thereby confirms and extends previous results \cite{Schon:2000qy, Boehmer:2008uq, Boehmer:2009sw, Thies:2018qgx,Thies:2022tpz}.
		The analysis is referred to as the stability analysis of the symmetric phase and does not require knowledge about spatial modulations of condensates.
		We perform this analysis in the infinite-$N$ limit at nonzero temperature and nonzero quark and chiral chemical potentials also inside the inhomogeneous phase.
		Thereby we observe an interesting relation between the bosonic $1$-particle irreducible two-point vertex function of the chiral Gross-Neveu model and the spinodal line of the Gross-Neveu model.
	\end{abstract}


	\maketitle

	\tableofcontents


\section{Introduction}

	In this work, we present another test of the applicability of a technique in \gls{qft} used to determine the phase structure in systems with condensation of order parameters: the so-called stability analysis of the ground state.
	Thereby, one typically examines the stability of either the phase without condensation or the phase with spatially homogeneous condensation \gls{wrt} (in)homogeneous perturbations.
	The method is particularily useful in searching for phases where the order parameter depends on the spatial coordinates -- a so-called \gls{ip} -- since it does not require a specific ansatz for the inhomogeneity.
	It is based on the determination of the bosonic $1$-particle-irreducible two-point function\footnote{In \gls{qft} textbooks, this object is defined as the second functional derivative \gls{wrt} the fields. The bosonic $1$-particle-irreducible two-point vertex function of a bosonic field $\phi$ is the inverse of $\langle \phi(x) \phi(y) \rangle$.}, abbreviated as bosonic two-point function in the following, and can be used in a variety of models, frameworks, and approximations, such as, e.g., the mean-field approximation or the \gls{frg}.
	It goes beyond a Ginzuburg-Landau analysis, since the momentum structure of the bosonic two-point function is completely resolved in contrast to an expansion in bosonic momenta.

	In this work, we apply and test the method with the $(1 + 1)$-dimensional \gls{chgn} model \cite{Gross:1974jv} in the infinite-$N$ limit at nonzero temperature and nonzero quark and chiral chemical potentials.
	
\subsection{Contextualization}

	In recent years, the stability analysis has been applied to a whole range of \gls{njl}-type models where it has been used to study inhomogeneous chiral order parameters, see Ref.~\cite{Nakano:2004cd,Abuki:2011pf,deForcrand:2006zz,Tripolt:2017zgc,Buballa:2020nsi,Buballa:2020xaa,Pannullo:2021edr,Pannullo:2022eqh,Pannullo:2023one,Pannullo:2023cat,Koenigstein:2023yzv,Winstel:2024dqu}.
	This work can be seen as a continuation of Refs.~\cite{Braun:2014fga,Koenigstein:2021llr}, which studied the ability of the method to detect the \gls{ip} in the $(1 + 1)$-dimensional \gls{gn} model, which features one spatially inhomogeneous scalar order parameter breaking translational invariance in addition to the discrete chiral symmetry spontaneously.
	In Refs.~\cite{Koenigstein:2021llr,Buballa:2014tba} one can (also) find an in-depth introduction to inhomgoeneous condensates in general, such that we abstain from repeating it here.
	In general, the stability analysis as it is formulated in this work can be seen as a slighly more modern version or extension of the \gls{adpt}.
	The latter was already applied to the \gls{chgn} model to detect the phase boundary \cite{Boehmer:2008uq,Boehmer:2009sw,Thies:2019ejd,Thies:2022tpz}.
	We therefore do not claim to bring up completely new findings or an entirely new approach, but rather present and test the stability analysis in a way that is more accessible to a broader audience and simpler to transfer to methods beyond mean field theory with minimal effort.

\subsection{Recap of existing results}

	The $(1 + 1)$-dimensional \gls{chgn} model is a generalization of the \gls{gn} model and features a continuous chiral symmetry and spatially inhomogeneous condensation of two bosonic fields breaking the continuous chiral symmetry as well as translational invariance.
	\begin{figure}
		\centering
		\includegraphics{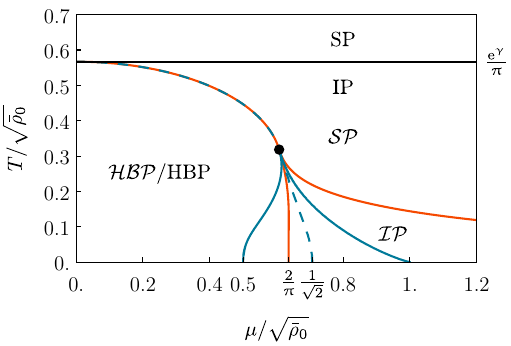}
		\caption{\label{fig:phasediagram}%
			Phase diagram of the $(1 + 1)$-dimensional \gls{chgn} model (black solid line and block letter labels) and \gls{gn} model (orange solid line and calligraphic labels).
			The blue dashed line is the phase diagram of both models under the assumption of a spatially homogeneous ground state.
			Blue solid lines are the inner and outer spinodal lines.
			The \glsfirst{hbp} for the \gls{chgn} model only exists under the assumption of a spatially homogeneous ground state.
			Otherwise the entire region below $T_\mathrm{c} = \frac{\ee^\upgamma}{\uppi}$ is an \gls{ip}.
			The phase diagrams are symmetric \gls{wrt} $\mu \to - \mu$ and constant in $\muchiral$-direction.
			The plot is a combination of results of Refs.~\cite{Sarma:1963,Jacobs:1974ys,Harrington:1974te,Harrington:1974tf,Dashen:1974xz,Schon:2000he,Schon:2000qy,Wolff:1985av,Schnetz:2004vr,Schnetz:2005ih,Schnetz:2005vh,Basar:2009fg,Boehmer:2007ea,Boehmer:2008uq,Boehmer:2009ae,Boehmer:2009sw,Thies:2018qgx,Thies:2019ejd}.
		}
	\end{figure}
	Within the mean-field (or infinite-$N$) approximation, the phase diagram of the $( 1 + 1 )$-dimensional \gls{chgn} model in the $( \mu, T )$ plane is rather simple, compare \cref{fig:phasediagram}.
	At the critical temperature $T_\mathrm{c} = \ee^\upgamma / \uppi$, it undergoes a second-order phase transition from the \gls{sp} without condensation to an \gls{ip} where two bosonic fields $\sigma$ and $\eta$ condense and form a so-called chiral spiral \cite{Schon:2000he,Schon:2000qy,Basar:2009fg,Boehmer:2007ea,Boehmer:2008uq,Boehmer:2009ae,Boehmer:2009sw,Thies:2018qgx,Thies:2019ejd,Ciccone:2023pdk,Ciccone:2023pdk}.
	As first obtained in Ref.~\cite{Schon:2000he} using an ansatz calculation, this chiral spiral is the solution for the thermodynamic ground state in the whole $( \mu, T )$ plane.
	It is given by
		\begin{align}
			&	\langle \sigma \rangle = m \cos 2 \mu (x - x_0) \, ,	&& \langle \eta \rangle = m\sin 2 \mu (x - x_0) \, ,	\label{eq:chiral_spiral}
		\end{align}	
	i.e., the two order parameters oscillate with a frequency determined by the chemical potential. The amplitude $m$ is a function of $T$ and goes to zero when $T$ approaches $T_\mathrm{c}$ from below.
	In Ref.~\cite{Thies:2018qgx}, it was further shown that the introduction of a chiral chemical potential $\muchiral$ does not change the behavior of the chiral order parameter such that the phase diagram is independent of $\muchiral$.

	As discussed above, the \gls{chgn} model can be understood as a generalization of the \gls{gn} model as it just contains one more interaction channel (mediated by $\eta$\footnote{In some publications it is called $\pi$.} in the bosonic picture).
	Through the continuous chiral symmetry transformation, one can show that the phase structure of the \gls{gn} and \gls{chgn} model are identical when restricting to homogeneous condensation.
	The homogeneous phase diagram of both models consists of a first order phase transition at low temperature, a critical point and a second-order phase transition at low chemical potential, see also \cref{fig:phasediagram}.
	However, the solution of the phase diagram when allowing for inhomogeneous condensation is vastly different, as can be seen from \cref{fig:phasediagram} when comparing the calligraphic and block letter labels for the phases.
	The \gls{chgn} model phase diagram consists only of a second-order phase transition to the \gls{ip} at $T=T_\mathrm{c}$ and $\vert\mu\vert > 0 $. This can be seen from \cref{eq:chiral_spiral} as $\hat{\sigma}$ and $\hat{\eta}$ are non-vanishing for $T < T_\mathrm{c}$. 
	In contrast, the \gls{gn} model phase diagram still features an \gls{hbp} at non-vanishing, low chemical potential and an \gls{ip} at temperatures $T$, which are significantly lower than $T_\mathrm{c}$, and high chemical potential.
	This difference is caused by the larger chiral symmetry group of the \gls{chgn} and the invariance of the chiral spiral under a combined, continuous translational and chiral transformation.
	An analogous invariance cannot be realized in the \gls{gn} model given the discrete chiral symmetry. 
	As we will discuss later, this feature of the \gls{chgn} model will also become useful in the context of the stability analysis.

	Several recent works are also tackling the problem of going beyond the mean-field approximation in order to incorporate bosonic fluctuations.
	While in the \gls{gn} model the literature is not fully conclusive regarding whether the existence of (quasi)-long range order at $T \neq 0$ somewhere in the phase diagram \cite{Lenz:2020bxk,Lenz:2020cuv,Stoll:2021ori}, recent results in \gls{chgn} model indicate that the chiral spiral solution is indeed destabilized by bosonic quantum fluctuations  such that only remnants of the \gls{ip} can be detected in bosonic correlation functions, which are oscillating but exponentially suppressed  \cite{Lenz:2021kzo,Lenz:2021vdz,Horie:2021wnn,Ciccone:2022zkg,Ciccone:2023pdk}.\footnote{%
	Regimes with oscillating but exponentially suppressed correlation functions in the phase diagram were recently termed ``quantum pion liquid'' (in analogy to ``quantum spin liquids in condensed matter systems) in higher-dimensional models and can be reached through a so-called disorder line \cite{Pisarski:2020dnx}.
	These regimes are related to the appearance of complex-conjugate poles with non-vanishing real and imaginary part in bosonic propagators \cite{Nishimura:2014kla,Schindler:2019ugo,Schindler:2021otf,Schindler:2021cke} and can also be obtained through disordering of inhomogeneous condensates by Goldstone modes of global $O(N)$ symmetry breaking \cite{Pisarski:2020dnx, Winstel:2024qle}.
	This has recently been observed in higher-dimensional fermionic models \cite{Haensch:2023sig, Winstel:2024dqu}.}

\subsection{Research objective}

	In this work, we apply the stability analysis to the $(1 + 1)$-dimensional \gls{chgn} model using the mean-field approximation, significantly simplifying the computations by neglecting bosonic quantum fluctuations.
	With this manuscript we particularly do not intend to bring up new findings regarding the phase structure of the \gls{chgn} model in the mean-field approximation, which was already derived in Refs.~\cite{Schon:2000he,Schon:2000qy,Basar:2009fg,Boehmer:2007ea,Boehmer:2008uq,Boehmer:2009ae,Boehmer:2009sw,Thies:2018qgx,Thies:2019ejd,Ciccone:2022zkg,Ciccone:2023pdk} and will be discussed below. 
	Instead, this manuscript should be understood as a further test of the applicability of the stability analysis in models where multiple order parameters form an inhomogeneous condensate.
	In this case, the corresponding stability analysis can be complicated by cross-correlations of these order parameters and diagonalization techniques are required in contrast to the simpler case of one order parameter as in the \gls{gn} model.
	A particular motivation for this work comes from Ref.~\cite{Winstel:2024dqu}, where the cross-correlations of the bosonic fields make an analytical treatment of the stability analysis impossible.
	Thus, the \gls{chgn} model serves as a good test case for applying this method to the condensation of multiple fields, since its phase diagram features inhomogeneous condensation of two order parameters, the solution to its phase diagram is well-known and rather simple.
	Furthermore, it extends previous stability anlysis in the form of the \gls{adpt} by resolving the full momentum structure of the bosonic two-point function also inside the inhomogeneous phase.

	In addition, we have realized that the stability analysis gives new insights of the relation between the inner spinodal line of the \gls{gn} and \gls{chgn} model and the momentum of the inhomogeneous order parameter in the \gls{ip} of the \gls{chgn} model.

	This work should therefore be seen as a guidance of how to set up a stability analysis in models with multiple fields with any method, where the momentum structure of the bosonic two-point function is accessible.

\subsection{Structure}

	This work is structured as follows.
	In \cref{sec:chiral_gn_model}, we introduce the $(1 + 1)$-dimensional \gls{chgn} model and discuss its symmetries, conserved charges, and chemical potentials.
	We then bosonize the theory, introduce a chiral basis in the bosonic field space, and present the expressions for the bare fermionic propagator and the fermion-boson vertices.
	In \cref{sec:bosonic_twopoint}, we derive the bosonic two-point function in the infinite-$N$ limit at nonzero temperature and nonzero quark and chiral chemical potentials.
	We then discuss the stability analysis of the symmetric phase and the detection of the phase boundary between the phase with chiral symmetry and the phase of spatially inhomogeneous chiral symmetry breaking in \cref{sec:results}.
	Finally, we conclude in \cref{sec:conclusions}.


\section{The chiral Gross-Neveu model}
\label{sec:chiral_gn_model}
	In this section, we define the $(1 + 1)$-dimensional \gls{chgn} model and introduce thermodynamic parameters such as non-zero temperature as well as quark and chiral chemical potentials in  the theory.
	After partial bosonization using a Hubbard-Stratonovich transformation and changing to a chiral basis in the bosonic field variables we present the relevant expressions for the fermionic propagator and fermion-boson vertices.

\subsection{Action in Minkowski space}

	We start with the action of the \gls{chgn} model in Minkowski space.
	Considering a $(1 + 1)$-dimensional spacetime, the action reads
		\begin{align}
			\S = \int \dd^2 x \, \Big[ \barpsi \, \ii \slashed{\partial} \, \psi + \tfrac{\ffcoupling}{2 N} \big[ ( \barpsi \, \psi )^2 + ( \barpsi \, \ii \gammachiral \, \psi )^2 \big] \Big] \, .	\label{eq:action_minkowski_vacuum}
		\end{align}
	Here, $\psi$ and $\barpsi$ are the fermionic fields, $\ffcoupling$ is the four-fermion coupling, and $N$ is the number of fermion flavors.
	We do not explicitly indicate the summation over the fermion flavors\footnote{The number of fermions in this model is sometimes, depending on the context, denoted as flavor or color in the literature.} and the Dirac indices.
	The chiral gamma matrix, which is the two-dimensional analog of the $\gamma^5$ from four-dimensional Dirac theory, is denoted by $\gammachiral$ and defined in \cref{eq:gamma_chiral_mink} in \cref{app:gamma_matrices}.
	The integration is over the two-dimensional spacetime coordinates $x^\mu = ( t, x )$, where $t \in ( - \infty, \infty )$ is the time coordinate and $x \in ( - \infty, \infty )$ is the spatial coordinate.
	We use the metric convention $\eta_{\mu \nu} = \mathrm{diag} ( + 1, - 1 )$.
	Whenever integration boundaries are not explicitly stated, we assume that the integration is performed over $\mathbb{R}$ from $- \infty$ to $+ \infty$ for each integral.

\subsection{Symmetries, conserved charges, and chemical potentials}

	Apart from spacetime and discrete symmetries, which shall not be discussed here, the \gls{chgn} model possesses two global continuous symmetries.
	The first symmetry is the $\U_\mathrm{V} ( 1 )$ symmetry (a global phase transformation)
	which leaves the four-fermion interaction terms separately invariant.
	The corresponding conserved current is the vector current $J^\mu_\mathrm{V} = \barpsi \, \gamma^\mu \, \psi \, ,$
	with the fermion number
		\begin{align}
			Q_\mathrm{V} = \int \dd^2 x \, J^0_\mathrm{V} = \int \dd^2 x \, \barpsi \, \gamma^0 \psi \, .
		\end{align}
	as conserved charge.
	The second symmetry is a $\U_\mathrm{A} ( 1 )$ symmetry, which acts on the fermionic fields as
		\begin{align}
			&	\psi \mapsto \psi^\prime = \ee^{\ii \vartheta \gammachiral} \, \psi \, ,	&&	\barpsi \mapsto \barpsi^\prime = \barpsi \, \ee^{\ii \vartheta \gammachiral} \, .
		\end{align}
	Note the appearance of the chiral gamma matrix $\gammachiral$ in the transformation as well as the sign in the transformation of the $\barpsi$ field.
	\Gls{wrt} these transformations the four-fermion interactions are no longer separately invariant, but they are transformed into each other.
	This also explains that they need to have the same coupling constant in order to preserve the symmetry.
	Only the discrete $\mathbb{Z}_2$ subgroup with $\vartheta = \tfrac{\uppi}{2} \, ( 2 n + 1 )$, $n \in \mathbb{Z}$ is a symmetry of each of the four-fermion interaction terms separately.
	Indeed, this is the symmetry of the \gls{gn} model which only has the scalar four-fermion interaction term.
	One identifies the conserved current of the $\U_\mathrm{A} ( 1 )$ symmetry with the axial current
		\begin{align}
			J^\mu_\mathrm{A} = \barpsi \, \gamma^\mu \, \gammachiral \, \psi \, ,
		\end{align}
	and the conserved charge as the axial charge
		\begin{align}
			Q_\mathrm{A} = \, & \int \dd^2 x \, J^0_\mathrm{A} =	\Vdistance
			\\
			= \, & \int \dd^2 x \, \barpsi \, \gamma^0 \, \gammachiral \psi = \int \dd^2 x \, \barpsi \, \gamma^1 \psi \, .	\Vdistance	\nonumber
		\end{align}
	To allow for an imbalance in the matter over anti-matter, we introduce a quark chemical potential $\mu$ in the microscopic action \labelcref{eq:action_minkowski_vacuum}.
	In addition, we introduce a chiral chemical potential $\muchiral$ to allow for a chiral imbalance.
		\begin{align}
			\S = \, & \int \dd^2 x \, \Big[ \barpsi \, ( \ii \slashed{\partial} + \mu \, \gamma^0 + \muchiral \, \gamma^1 ) \, \psi +	\Vdistance
			\\
			& \quad + \tfrac{\ffcoupling}{2 N} \big[ ( \barpsi \, \psi )^2 + ( \barpsi \, \ii \gammachiral \, \psi )^2 \big] \Big] \, .	\Vdistance	\nonumber
		\end{align}
	Note that the sign in front of the chemical potentials depends on the convention for what we call matter and anti-matter etc.\ and is of no importance for the following discussion.

	Furthermore, note that the $U_\mathrm{A} ( 1 )$ symmetry only exists on the classical level but is broken by the path integral measure in terms of an anomaly.
	For $N \to \infty$ we can ignore its breaking and treat it as a symmetry of the model, while its role at finite $N$ was recently discussed in Ref.~\cite{Ciccone:2022zkg,Ciccone:2023pdk}.

\subsection{Euclidean action}

	Within this work we are interested in the statistical properties of the \gls{chgn} model, which is why we switch to Euclidean spacetime.
	This also allows us to study the model at nonzero temperature.
	We perform the Wick rotation $t \to - \ii \tau$ to obtain the Euclidean action.
	In addition to the temporal coordinate, we also Wick rotate the spatial gamma matrix $\gamma^1 \to - \ii \gamma^1$.
	The resulting Euclidean action is
		\begin{align}
			\S = \, & \int \dd^2 x \, \Big[ \barpsi \, ( \slashed{\partial} - \mu \, \gamma^0 + \ii \muchiral \, \gamma^1 ) \, \psi +	\Vdistance
			\\
			& \quad - \tfrac{\ffcoupling}{2 N} \big[ ( \barpsi \, \psi )^2 + ( \barpsi \, \ii \gammachiral \, \psi )^2 \big] \Big] \, .	\Vdistance	\nonumber
		\end{align}
	Note the sign change in front of the four-fermion interaction term \gls{wrt} the kinetic term as well as the additional factor $\ii$ in front of the chiral chemical potential.

\subsection{Partial bosonization}

	Before we introduce temperature to the problem, we perform a Hubbard-Stratonovich transformation to (partially) bosonize the theory.
	The resulting action reads
		\begin{align}
			\S = \, & \int \dd^2 x \, \Big[ \tfrac{\ycoupling^2}{2 \ffcoupling} \, ( \phi^2 + \xi^2 ) +	\Vdistance
			\\
			& \quad + \barpsi \, \big( \slashed{\partial} - \mu \, \gamma^0 + \ii \muchiral \, \gamma^1 + \tfrac{\ycoupling}{\sqrt{N}} \, ( \phi + \ii \gammachiral \, \xi ) \big) \, \psi \Big] \, ,	\Vdistance	\nonumber
		\end{align}
	where the four-fermion interaction in the scalar channel is replaced by a Yukawa coupling $\ycoupling$ of the fermions to bosonic $\phi$ field, while the four-fermion interaction in the pseudo-scalar channel is replaced by a Yukawa coupling of the fermions to a bosonic $\xi$ field.
	Note that the transformation is exact such that partition functions and correlation functions are equivalent and can either be expressed in terms of the fermionic or the bosonic fields.
	Concerning the $U_\mathrm{V} ( 1 )$ symmetry, the bosonic fields do not transform at all.
	However, the $U_\mathrm{A} ( 1 )$ symmetry rotates the bosonic fields into each other \cite{Gross:1974jv},
		\begin{align}
			\begin{pmatrix}
				\phi
				\\
				\xi
			\end{pmatrix}
			\mapsto
			\begin{pmatrix}
				\phi^\prime
				\\
				\xi^\prime
			\end{pmatrix}
			=
			\begin{pmatrix}
				\cos ( 2 \vartheta ) & \sin ( 2 \vartheta )
				\\
				- \sin ( 2 \vartheta ) & \cos ( 2 \vartheta )
			\end{pmatrix}
			\begin{pmatrix}
				\phi
				\\
				\xi
			\end{pmatrix} \, ,
		\end{align}
	which corresponds to an $O ( 2 )$ symmetry -- the symmetry of the effective bosonic potential that we are going to study later.
	
	Since we will be working in the $N \to \infty$ limit, it is convenient to absorb the $\ycoupling$ and a factor $\frac{1}{\sqrt{N}}$ into the bosonic fields,
		\begin{align}
			&	\sigma = \tfrac{\ycoupling}{\sqrt{N}} \, \phi \, ,	&&	\eta = \tfrac{\ycoupling}{\sqrt{N}} \, \xi \, .
		\end{align}
	This is possible, because there is no loop correction to the Yukawa coupling for $N \to \infty$.
	However, note that this changes the dimensions of the bosonic fields, which are now of dimension energy.

\subsection{Temperature}

	Finally, we introduce temperature in the usual way by compactifying the Euclidean time direction to a circle of circumference $\beta = \tfrac{1}{T}$.
	Fermions are supposed to obey anti-periodic boundary conditions in the temporal direction, while bosons are supposed to obey periodic boundary conditions.
	The resulting action reads
		\begin{align}
			\S = \, & \int_{0}^{\frac{1}{T}} \dd \tau \int \dd x \, \Big[ \tfrac{N}{2 \ffcoupling} \, ( \sigma^2 + \eta^2 ) +	\Vdistance
			\\
			& \quad + \barpsi \, \big( \slashed{\partial} - \mu \, \gamma^0 + \ii \muchiral \, \gamma^1 + \sigma + \ii \gammachiral \, \eta \big) \, \psi \Big] \, .	\Vdistance	\nonumber
		\end{align}
	A direct consequence of the compactification of the temporal direction is the quantization of the bosonic and fermionic energies in terms of the Matsubara frequencies $\nu_n = 2 \uppi T ( n + \frac{1}{2} )$ for fermions and $\omega_n = 2 \uppi T n$ for bosons with $n \in \mathbb{Z}$.
	Conventions for the Fourier transformations to momentum space are taken from the Appendix of Ref.~\cite{Koenigstein:2023yzv}.

\subsection{A chiral basis in field space}

	When studying the bosonic two-point functions of the \gls{chgn} model in the basis of $\sigma$ and $\eta$ fields directly, off-diagonal contributions in the bosonic two-point function are obtained and diagonalization is required\footnote{Interesting physical properties are anyhow contained in the eigenvalues.
	References \cite{Basar:2009fg,Boehmer:2007ea,Boehmer:2008uq,Boehmer:2009ae,Boehmer:2009sw,Thies:2018qgx,Thies:2019ejd} indeed study the determinant within the \gls{adpt} approach.}.
	Hereby, we found that the basis in field space, where the bosonic two-point function is diagonal, is a basis that we will call the chiral basis.
	It was for example already introduced in Refs.~\cite{Basar:2008im,Basar:2008ki,Basar:2009fg,Dunne:2013xta,Thies:2018qgx,Lenz:2021kzo} and is given by\footnote{Note that we did not normalize the transformation matrix.
	This is not necessary and we would only complicate several expressions by additional factors of $\frac{1}{\sqrt{2}}$.}
		\begin{align}
			&	\varphileft = \sigma - \ii \eta \, ,	&&	\varphiright = \sigma + \ii \eta \, ,	\vdistance	\label{eq:chiral_basis_1}
			\\
			&	\sigma = \tfrac{1}{2} \, ( \varphileft + \varphiright ) \, ,	&&	\ii \eta = \tfrac{1}{2} \, ( \varphiright - \varphileft ) \, ,	\vdistance	\label{eq:chiral_basis_2}
		\end{align}
	and one finds with \cref{eq:chiral_projectors_euclidean} that
		\begin{align}
			\sigma \, \mathds{1} + \ii \gammachiral \, \eta = \gammaleft \, \varphileft + \gammaright \, \varphiright \, .
		\end{align}
	Hence, the microscopic action that we are working with is given by,
		\begin{align}
			\S = \, & \int_{0}^{\frac{1}{T}} \dd \tau \int \dd x \, \Big[ \tfrac{N}{2 \ffcoupling} \, \varphileft \, \varphiright +	\Vdistance	\label{eq:action_final}
			\\
			& \quad + \barpsi \, \big( \slashed{\partial} - \mu \, \gamma^0 + \ii \muchiral \, \gamma^1 + \gammaleft \, \varphileft + \gammaright \, \varphiright  ) \, \psi \Big] \, .	\Vdistance	\nonumber
		\end{align}

\subsection{The two-point function, the propagator, and vertices}
	For the latter computation of the effective bosonic potential and the bosonic two-point function, the expressions for the bare fermionic two-point function and propagator as well as the fermion-boson vertices are required.
	These are obtained by transforming the action \labelcref{eq:action_final} to Fourier momentum/frequency space with the conventions from Ref.~\cite{Koenigstein:2023yzv} and taking derivatives \gls{wrt} the fermionic fields.
	The bare fermionic two-point function evaluated for constant bosonic background fields reads
		\begin{align}
			& S^{\barpsi_2 \psi_1} ( \nu_{n_2}, p_2; \nu_{n_1}, p_1) =	\vdistance	\label{eq:fermionic_twopoint}
			\\
			= \, & - \beta \, \delta_{n_2, n_1} \, 2 \uppi \, \delta ( p_2 - p_1 ) \times	\vdistance	\nonumber
			\\
			& \quad \times \big[ \ii ( \nu_{n_1} + \ii \mu ) \, \gamma^0 + \ii ( p_1 + \muchiral ) \, \gamma^1 + \gammaleft \, \varphileft + \gammaright \, \varphiright \big] \, .	\vdistance	\nonumber
		\end{align}
	Here, we only used left derivatives and derivatives are taken in the order of the fields in the superscript (right to left).
	The fermionic propagator for constant bosonic background fields is the inverse of the two-point function,
		\begin{align}
			& ( S^{\barpsi_2 \psi_1} )^{- 1} ( \nu_{n_2}, p_2; \nu_{n_1}, p_1) =	\vdistance	\label{eq:fermionic_propagator}
			\\
			= \, & \beta \, \delta_{n_2, n_1} \, 2 \uppi \, \delta ( p_2 - p_1 ) \times	\vdistance	\nonumber
			\\
			& \quad \times \frac{\ii ( \nu_{n_1} + \ii \mu ) \, \gamma^0 + \ii ( p_1 + \muchiral ) \, \gamma^1 - \gammaleft \, \varphiright - \gammaright \, \varphileft}{( \nu_{n_1} + \ii \mu )^2 + ( p_1 + \muchiral )^2 + \varphileft \, \varphiright} \, .	\vdistance	\nonumber
		\end{align}
	The only other ingredient that we need are the fermion-boson vertices, which are given by
		\begin{align}
			& S^{\varphileftright_3 \barpsi_2 \psi_1} ( \omega_{n_3}, p_3; \nu_{n_2}, p_2; \nu_{n_1}, p_1) =	\vdistance	\label{eq:fermion_boson_vertices}
			\\
			= \, & - \beta \, \delta_{n_2, n_3 + n_1} \, 2 \uppi \, \delta ( p_3 - p_2 + p_1 ) \, \gammaleftright \, .	\vdistance	\nonumber
		\end{align}

\section{Calculation of the bosonic two-point function -- formal setup for the stability analysis}
\label{sec:bosonic_twopoint}

	In this section, we derive an expression for the bosonic two-point function of the \gls{chgn} model in the infinite-$N$ limit.
	We start by discussing the renormalization procedure and calculating the renormalized effective potential under the assumption of a homogeneous background field configuration.
	In fact, this step is not really necessary for the stability analysis, if one is exclusively interested in second-order phase transitions between the \gls{sp} and a \gls{hbp} or the \gls{sp} and an \gls{ip}, but is, however, a nice consistency check\footnote{
	The reason is that one studies the stability of the symmetric phase by evaluating the bosonic two-point function at vanishing background field configuration, which does not require knowledge of the position of the minimum of the effective potential in a \gls{hbp}.
	One basically stops the analysis once the phase transition is found and actually never enters phases of symmetry breaking.}.
	Finally, we calculate the bosonic two-point function in the infinite-$N$ limit (for vanishing bosonic background field) and discuss the result.

\subsection{Effective potential}

	The starting point is the effective action of the \gls{chgn} model.
	Since we work in the infinite-$N$ limit, the effective action $\Gamma$ is given by the $\frac{1}{N}$-rescaled exponent in the partition function after integrating out the fermions (the zeroth order of a saddle-point expansion).
	Additionally, one has to divide by the volume and inverse temperature of the system.
		\begin{align}
			& ( \beta V ) \, \Gamma = \tfrac{1}{N} \, \S =	\Vdistance	\label{eq:effective_action}
			\\
			= \, & \int \dd x \int_{0}^{\frac{1}{T}} \dd \tau \, \Big[ \tfrac{1}{2 \ffcoupling} \, \varphileft \, \varphiright +	\Vdistance	\nonumber
			\\
			& - \ln \Det \big[ \beta \, \big( \slashed{\partial} - \mu \, \gamma^0 + \ii \muchiral \, \gamma^1 + \gammaleft \, \varphileft + \gammaright \, \varphiright  ) \big] \, .	\Vdistance	\nonumber
		\end{align}
	Another way of thinking of this is to study the one-loop correction to the classical bosonic action, hence including only fermion fluctuations (which is the leading order in the $1 / N$-expansion).
	Since we do not approach the problem of \glspl{ip} directly, but via the stability of spatially homogeneous phases, we will only consider constant background field configurations for the bosonic fields, which is indicated by bar symbols.
	Evaluating $\Gamma$ for constant background fields, we obtain the effective potential $\bar{U}$,
		\begin{align}
			\bar{U} = \, & \tfrac{1}{2 \ffcoupling} \, \barvarphileft \, \barvarphiright +	\Vdistance
			\\
			& - \tfrac{1}{\beta V} \, \ln \Det \big[ \beta \, \big( \slashed{\partial} - \mu \, \gamma^0 + \ii \muchiral \, \gamma^1 + \gammaleft \, \barvarphileft + \gammaright \, \barvarphiright  ) \big] \, .	\Vdistance	\nonumber
		\end{align}
	The steps leading to the final result are standard, can be found elsewhere, e.g., Refs.~\cite{Koenigstein:2021llr,Koenigstein:2023yzv,Koenigstein:2023wso,Pannullo:2023cat}, and we will only summarize them briefly.
	\begin{enumerate}
		\item	We introduce the $O(2)$-invariant of the background fields $\bar{\rho} = \barvarphileft \, \barvarphiright = \sigma^2 + \eta^2$.
		
		\item	We switch to momentum/frequency space and insert the bare fermionic two-point function \labelcref{eq:fermionic_twopoint} in the determinant, which can be evaluated in Dirac space using the explicit representation of the gamma matrices from \cref{app:gamma_matrices}.
		
		\item	Afterwards we use $\ln \Det = \Tr \ln$.
		The remaining (functional) trace is the sum over Matsubara frequencies and a spatial momentum integral.

		\item	The Matsubara sum is standard, see for example Refs.~\cite{LeBellac:1991cq, Kapusta:2006pm}.
		Note, that it is only defined up to an infinite constant, which is irrelevant for the following discussion.
	\end{enumerate}
	The form of the result is well-known and reads
		\begin{align}
			\bar{U} = \, & \tfrac{1}{2 \ffcoupling} \, \bar{\rho} - \int_{- \infty}^{\infty} \frac{\dd p}{2 \uppi} \, \big( \tilde{E} + \tfrac{1}{\beta} \big[ \ln \big( 1 + \ee^{- \beta ( \tilde{E} + \mu )}  \big) +	\Vdistance	\label{eq:effective_potential}
			\\
			& \qquad + ( \mu \to - \mu ) \big] \, ,	\Vdistance	\nonumber
		\end{align}
	where we defined the energy of the fermions (in the presence of $\muchiral$) as
		\begin{align}
			& E = \sqrt{p^2 + \bar{\rho}} \, ,	&&	\Leftrightarrow	&&	\tilde{E} = \sqrt{( p + \muchiral )^2 + \bar{\rho}} \, .
		\end{align}
	Here, we can shift the momentum integral by $\muchiral$ such that the effective potential does not depend on $\muchiral$ except for an overall constant term, which is derived in the context of renormalization below.
	Some remarks are in order.
	\begin{enumerate}
		\item	This is the standard \gls{gn} result for the effective potential, where $\sigma^2$ is replaced by $\bar{\rho} = \sigma^2 + \eta^2$.
		Hence, the homogeneous phase diagram is identical to the one of the \gls{gn} model, which is well-known \cite{Schon:2000he,Schon:2000qy,Basar:2009fg,Boehmer:2007ea,Boehmer:2008uq,Boehmer:2009ae,Boehmer:2009sw,Thies:2018qgx,Thies:2019ejd}.
		
		\item	The $O ( 2 )$-symmetry of the potential is manifest, because it exclusively depends on the invariant $\bar{\rho}$.
		This is a results of the $U_\mathrm{A} (1)$-symmetry, hence chiral symmetry is realized as a continuous symmetry also on the bosonic level.
		However, a ground state of the system with a nonzero expectation value for $\bar{\rho}$ as it is realized in the \gls{hbp} for certain $\mu$ and $T$ spontaneously breaks the $O(2)$-symmetry and therefore also the $U_\mathrm{A} (1)$ (chiral) symmetry.

		\item	The bosonic potential does (appart from an overall constant) not depend on $\muchiral$, which implies that the homogeneous phase diagram of the \gls{chgn} model is independent of the chiral chemical potential, which is a known result.
		As we will see and as it is also known, this also holds true for the full phase diagram with the \gls{ip}.
		However, note that if the model is for example studied in a finite sized box or on a lattice, the dependence on $\muchiral$ via the momentum integral is more complicated and the chiral chemical potential can have an impact on the phase diagram.
	\end{enumerate}

\subsection{The gap equation and renormalization of the effective potential}

	The effective potential in the form of \cref{eq:effective_potential} is acually not finite (because of the first term in the integrand) and still depends on the four-fermion coupling.
	To renormalize the theory, we have to subtract the divergent part of the effective potential.
	One option to do so is to use the gap equation in vacuum as a renormalization condition:
	In vacuum one has symmetry breaking and finds degenerate nontrivial minima of the effective potential at $\bar{\rho}_0$.
	The condition for these minima is
		\begin{align}
			0 \overset{!}{=} \tfrac{\dd \bar{U}}{\dd \bar{\rho}} \Big|_{\bar{\rho} = \bar{\rho}_0} = \tfrac{1}{2 \lambda} - \tfrac{1}{4 \uppi} \int_{- \infty}^{\infty} \dd p \, \tfrac{1}{\sqrt{p^2 + \bar{\rho}_0}} \, .
		\end{align}
	One discovers a logarithmic \gls{uv} divergence, which we regularize with a sharp cutoff $\Lambda$.
	Further, we absorb the divergence into the definition of the four-fermion coupling.
	Solving for $\frac{1}{\lambda}$ one finds
		\begin{align}
			\tfrac{1}{\lambda} = \tfrac{1}{2 \uppi} \ln \big( \tfrac{4 \Lambda^2}{\bar{\rho}_0} \big) + \mathcal{O} \big( \tfrac{\bar{\rho}_0}{\Lambda^2} \big) \, .
		\end{align}
	This is now used to renormalize the effective potential:
	Regularizing the divergent contribution of $\bar{U}$ with the same sharp \gls{uv} cutoff $\Lambda$ one finds (ignoring a field independent term $\propto \Lambda^2$)
		\begin{align}
			\int_{- \Lambda}^{\Lambda} \dd p \, \tfrac{\tilde{E}}{2 \uppi} = - \tfrac{1}{4 \uppi} \, \bar{\rho} \, \big[ \ln \big( \tfrac{\bar{\rho}}{4 \Lambda^2} \big) - 1 \big] + \tfrac{\muchiral^2}{2 \uppi} + \mathcal{O} \big( \tfrac{\bar{\rho}}{\Lambda^2} \big) \, .
		\end{align}
	Hence, inserting this and the regularized expression for $\frac{1}{\lambda}$ into the effective potential one can safely send $\Lambda \to \infty$, such that the final result is
		\begin{align}
			\bar{U} = \, & \tfrac{1}{4 \uppi} \, \bar{\rho} \, \big[ \ln \big( \tfrac{\bar{\rho}}{\bar{\rho}_0} \big) - 1 \big] - \tfrac{\muchiral^2}{2 \uppi} + 	\Vdistance
			\\
			& - \tfrac{1}{\uppi} \int_0^\infty \dd p \, \tfrac{1}{\beta} \big[ \ln \big( 1 + \ee^{- \beta ( E + \mu )}  \big) + ( \mu \to - \mu ) \big]	\Vdistance	\nonumber
		\end{align}
	Note that the effective potential is now (up to irrelevant diverging constants) finite and all dimensionful quantities can be expressed in units of $\bar{\rho}_0$, which sets the scale of the theory.
	This expression can now be used to determine the spatially homogeneous ground state and the homogeneous phase diagram, see \cref{fig:phasediagram}.
	Note, that the constant shift proportional to $\muchiral^2$ is irrelevant for the phase diagram, because it does not affect the position of the minimum.
	
\subsection{The bosonic two-point functions}

	Next, we turn to the computation of the bosonic two-point functions.
	We do not recapitule the entire construction principle behind the stability analysis at this point, see Ref.~\cite{Koenigstein:2021llr} for a detailed derivation of the stability analysis.
	Basically we take the effective action from \cref{eq:effective_action} and expand it to second order in the bosonic fields.\footnote{When performing the expansion, it turns out that the first order corrections are proportional to the gap equation and, thus, vanish when performing the expansion about extrema of the effective action.}
	Hence, we are studying the quadratic fluctuations around the homogeneous ground state.
	We are interested in the stability of the \gls{sp}, which is why we evaluate the two-point matrix elements directly at vanishing background field configuration.
	Thus, we are interested in the coefficient $\Gamma^{(2)} ( \mu, T, q )$ in the quadratic order of the expansion, which explicitly reads,
		\begin{align}
			\beta \int \tfrac{\dd q}{2 \uppi} \, ( \delta \sigma ( - q ), \delta \eta ( - q ) ) \cdot \Gamma^{(2)} ( \mu, T, q ) \cdot
			\begin{pmatrix}
				\delta \sigma ( q )
				\\
				\delta \eta ( q )
			\end{pmatrix} \, .	\label{eq:quadratic_coefficient_effective_action}
		\end{align}
	Formally, the only thing we have to do is to take two functional derivatives of the effective action \labelcref{eq:effective_action} \gls{wrt} the bosonic field-space vectors and evaluate the result at vanishing background field configuration and vanishing external Matsubara frequencies.
	The resulting expression depends on the external bosonic momentum and is a two-by-two matrix in field space whose eigenvalues are the two bosonic two-point functions of the system and determine the stability of the symmetric phase.
	(Eigenvalues that are negative --  here depending on $\mu$, $T$, and $q$ -- indicate an instability of the symmetric phase \gls{wrt} a fluctuation of momentum $q$.)
	
	As already explained above, we work in the chiral basis.
	The transformation matrices that can be read off from \cref{eq:chiral_basis_1,eq:chiral_basis_2} are given by
		\begin{align}
			&	M =
			\begin{pmatrix}
				1 & - \ii
				\\
				1 & \ii
			\end{pmatrix} \, ,
			&&	M^{- 1} = \tfrac{1}{2}
			\begin{pmatrix}
				1 & 1
				\\
				\ii & - \ii
			\end{pmatrix} \, .
		\end{align}
	Inserting $\mathds{1} = M \cdot M^{- 1} = M^{- 1} \, M$ in \cref{eq:quadratic_coefficient_effective_action} between the fields and the coefficient matrix, one finds
		\begin{align}
			\beta \int \tfrac{\dd q}{2 \uppi} \, 
			( \delta \varphiright ( - q ), \delta \varphileft ( - q ) ) \cdot \tilde{\Gamma}^{(2)} ( \mu, T, q ) \cdot
			\begin{pmatrix}
				\delta \varphileft ( q )
				\\
				\delta \varphiright ( q )
			\end{pmatrix} \, ,	\label{eq:quadratic_coefficient_effective_action_2}
		\end{align}
	with
		\begin{align}
			\tilde{\Gamma}^{(2)} ( \mu, T, q ) = M \, \Gamma^{(2)} ( \mu, T, q ) \, M^{- 1}
		\end{align}
	being the coefficient matrix in the chiral basis.
	This, however, can be directly calculated by taking functional derivatives of \cref{eq:effective_action} \gls{wrt} $\Phi = ( \varphileft, \varphiright )^T$ and $\Phi^\dagger = ( \varphiright, \varphileft )$ instead of $( \sigma, \eta )^T$ and $( \sigma, \eta )$, again evaluating the result at vanishing background field configuration and vanishing external bosonic Matsubara frequencies.
	
	The general unrenormalized expression for the coefficient matrix -- containing the bosonic two-point functions -- reads
		\begin{align}
			& \beta 2 \uppi \delta ( q_\mathrm{II} + q_\mathrm{I} ) \, \tilde{\Gamma}^{(2)}_{ji} ( \mu, T, q_\mathrm{I} ) =	\vdistance	\label{eq:unevaluated_bosonic_twopoint}
			\\
			= \, & \beta \, 2 \uppi \, \delta ( q_\mathrm{II} + q_\mathrm{I} ) \, \delta_{j i} \, \tfrac{1}{2 \ffcoupling} +	\vdistance	\nonumber
			\\
			& + \Tr \Big( ( \S^{(2)} )^{- 1} \, \S^{(3)}_{\Phi^\dagger_j ( 0, q_\mathrm{II} )} \, ( \S^{(2)} )^{- 1} \, \S^{(3)}_{\Phi_i ( 0, q_\mathrm{I} )} \Big)	\vdistance	\nonumber
		\end{align}
	Here, the indices $i$ and $j$ correspond to the entries of the field-space vectors $\Phi$ and $\Phi^\dagger$.
	Furthermore, $( \S^{(2)} )^{- 1}$ is a shorthand notation for fermionic propagator \labelcref{eq:fermionic_propagator} (at $\varphileft = \varphiright = 0$) and $\S^{(3)}$ for the fermion-boson vertices \labelcref{eq:fermion_boson_vertices}, where the fermionic indices are traced over.
	The trace includes sums over Matsubara frequencies and momentum integrals as well as a trace over Dirac matrices.
	Explicit expressions for the traces and their evaluations are presented in the \cref{app:evaluation_of_the_field_theory_traces}.
	One finds that the offdiagonal entries of the coefficient matrix vanish and the diagonal entries -- the bosonic two-point functions -- are given by
		\begin{align}
			& \tilde{\Gamma}^{(2)}_{11/22} ( \mu, T, q ) =	\Vdistance
			\\
			= \, & \tfrac{1}{2 \lambda} - \tfrac{\dimDirac}{2} \, \tfrac{1}{\beta} \sum_{n} \int \frac{\dd p}{2 \uppi} \, \frac{1}{\big[ \big( \nu_n + \ii \big( \mu \pm \tfrac{q}{2} \big) \big)^2 + p^2 \big]} =	\Vdistance	\nonumber \\
			= \, & \tfrac{1}{2 \lambda} - \int \frac{\dd p}{2 \uppi} \, \frac{1}{2 p} \, \Big[ 1 - \nf \Big( \tfrac{p + ( \mu \pm \frac{q}{2} )}{T} \Big) - \nf \Big( \tfrac{p - ( \mu \pm \frac{q}{2} )}{T} \Big) \Big] \, .	\Vdistance	\nonumber
		\end{align}
	Here, the the evaluation of the Matsubara sum is a well-known calculation \cite{LeBellac:1991cq, Koenigstein:2023yzv,Koenigstein:2023wso}.
	Again, we already shifted the spatial momentum by $\muchiral$, such that the final expression is independent of $\muchiral$.
	Of course, the momentum integral over the first integrand is not \gls{uv} finite.
	However, this divergence is canceled by the divergence of the $\frac{1}{\lambda}$ contribution.\footnote{Formally, the integral boundaries of the regularized vacuum integral are $\muchiral$- and $q$-dependent after the momentum shifts that were performed in \cref{app:evaluation_of_the_field_theory_traces}.
	However, after sending $\Lambda \to \infty$ during the this dependence vanishes.}
	The tricky part is acually the \gls{ir} divergence, which appears in every term under the integral.
	It turns out that these divergences cancel each other and the final result is \gls{uv} and \gls{ir} finite.
	We do not present the explicit calculations here and refer to the appendix of Ref.~\cite{Koenigstein:2023wso} for details.
	As the final result, we obtain
		\begin{align}
			& \tilde{\Gamma}^{(2)}_{11/22} ( \mu, T, q ) =	\vdistance	\label{eq:bosonic_twopoint_final_T}
			\\
			= \, & \tfrac{1}{2 \uppi} \Big[ \tfrac{1}{2} \ln \big( \tfrac{4 T^2}{\bar{\rho}_0} \big) - \mathrm{DLi}_0 \Big( \tfrac{\mu \pm \frac{q}{2}}{T} \Big) - \upgamma \Big]	\vdistance	\nonumber
		\end{align}
	for nonzero $T$, and
		\begin{align}
			\tilde{\Gamma}^{(2)}_{11/22} ( \mu, 0, q ) = \tfrac{1}{4 \uppi} \ln \Big( \tfrac{4 ( \mu \pm \frac{q}{2} )^2}{\bar{\rho}_0} \Big)	\vdistance	\label{eq:bosonic_twopoint_final_0}
		\end{align}
	for $T = 0$.\footnote{Note that an expansion of \cref{eq:bosonic_twopoint_final_T} about $q = 0$ has to agree with the Ginzburg-Landau coefficients of the \gls{chgn} model, see Ref.~\cite{Basar:2009fg,Boehmer:2007ea,Boehmer:2009sw}.}
	In the first expression we introduced
		\begin{align}
			& \mathrm{DLi}_{2n} ( y ) =	\vdistance
			\\
			= \, & \big[ \tfrac{\partial}{\partial s} \, \mathrm{Li}_s ( - \ee^y ) + \tfrac{\partial}{\partial s} \, \mathrm{Li}_s ( - \ee^{-y} )\big]_{s = 2n} =	\vdistance	\nonumber
			\\
			= \, & - \delta_{0,n} ( \ln ( 2 \uppi ) + \upgamma ) +	\vdistance	\nonumber
			\\
			& + ( - 1 )^{1-n} ( 2 \uppi )^{2n} \, \mathrm{Re} \, \psi^{(-2n)} \big( \tfrac{1}{2} + \tfrac{\ii}{2 \uppi} \, y \big) \, .	\vdistance	\nonumber
		\end{align}
	Here, $\mathrm{Li}_s$ is the polylogarithm function and $\psi^{(n)}$ the polygamma function, while $\upgamma$ is the Euler-Mascheroni constant.
	This result is in agreement with taking derivatives of the free energy from Ref.~\cite{Ciccone:2022zkg,Ciccone:2023pdk}.
	
	Connaisseurs of the \gls{gn} model will recognize the here presented bosonic two-point functions as the bosonic two-point function of the \gls{gn} model at vanishing background field configuration and vanishing external momentum, see Ref.~\cite[Eqs.~(A8) \& (A.16)]{Koenigstein:2021llr} or Ref.~\cite[Eq.~(F.37) \& (F.57)]{Koenigstein:2023wso}, though shifted chemical potential, $\mu \to \mu \pm \frac{q}{2}$.\footnote{Another difference is a global factor of $\frac{1}{2}$ which stems from the change to the chiral basis without normalized basis change matrices.}
	Hence, the external momentum no longer appears separately but solely as a shift of $\mu$.
	It is known that the bosonic two-point function of the \gls{gn} model for trivial background field configuration and zero external momentum vanishes (has its sign change) exactly on the inner spinodal line going over to the second order phase transition (the inner blue curve in \cref{fig:phasediagram}).
	Hence, there will be a direct connection between the inner spinodal line of the \gls{gn} model and the second order phase transition from the \gls{sp} to the \gls{ip} in the \gls{chgn} model.
	This is explained in the next section.

\section{Results}
\label{sec:results}

	We are now ready to present the results of this work.
	This comprises the stability analysis of the \gls{chgn} model in the infinite-$N$ limit and the momentum structure of the bosonic two-point functions, which in turn reproduces the known phase diagram of the \gls{chgn} model from \cref{fig:phasediagram}.
	Furthermore, we discuss a connection between the inner spinodal line of the \gls{gn} model and the second order phase transition from the \gls{sp} to the \gls{ip} in the \gls{chgn} model.

\subsection{Stability analysis}

	As already discussed in the introduction, the simple idea behind the stability analysis is to study the stability of the spatially homogeneous ground state instead of aiming for the full phase diagram as well as the exact shape of possible modulations within an \gls{ip}.
	As discussed in detail in previous works, see, e.g.\ Ref.~\cite{Koenigstein:2021llr}, this approach is sufficient as long as the phase transition is of second order and as long as one approaches the phase transition from the \gls{sp}.
	Otherwise, the assumption of fluctuations with an infinitesimal amplitude around a homogeneous ground state is not necessarily valid.
	In addition to detecting the phase transition to an \gls{ip} the method also allows to detect the phase transition to a \gls{hbp}.
	Here, one of the two bosonic two-point functions has vanishing eigenvalues at vanishing external momentum $q$.\footnote{At the phase transition, this agrees with the vanishing determinant condition of \gls{adpt} of Refs.~\cite{Boehmer:2008uq,Boehmer:2009sw,Thies:2019ejd,Thies:2022kuv}.}
	
	Thus, the strategy is as follows:
	One searches for the minimum of the effective potential for different combinations of $\muchiral$-$\mu$-$T$ and determines the phase diagram under the assumption of spatially homogeneous condensation.
	In regions without condensation, one studies the bosonic two-point functions \labelcref{eq:bosonic_twopoint_final_T}  as a function of the external bosonic momentum -- the momentum of the perturbations.
	If one finds eigenvalues that are negative at $q\neq 0$, the \gls{sp} is unstable and the system develops a spatially inhomogeneous condensate instead of a trivial ground state.
	However, it is not possible by this method to safely test the stability of a nonvanishing groundstate in the \gls{hbp}.
	
	In fact, one does not even need to determine the homogeneous phase diagram before starting the stability analysis.
	\begin{table}
		\renewcommand{\arraystretch}{1.3}
		\centering
		\caption{\label{tab:roots_eigenvalue}%
			The two roots $q_{0,i}$, $i \in \{ 1, 2\},$ of the second eigenvalue of the bosonic two-point function \labelcref{eq:bosonic_twopoint_final_T} for different $T < T_\mathrm{c}$ at fixed $\mu = \sqrt{\bar{\rho}_0}$, see also \cref{fig:gamma2ofq}, and the corresponding chemical potentials $\mu_\mathrm{spinodal}$ of the inner spinodal line and second order phase transition of the \gls{gn} model at the same $T$.
			One finds $\mu_\mathrm{spinodal} = \mu - \tfrac{q_{0,i}}{2}$.
			All values are in units of $\sqrt{\bar{\rho}_0}$ and floats are rounded to the third digit.%
		}%
		\begin{ruledtabular}
			\begin{tabular}{c c c c}
				$T$	&	$q_{0,1}$	&	$q_{0,2}$	&	$\mu_\mathrm{spinodal}$
				\\
				\hline
				$0$	&	$1$	&	$3$	&	$\pm \frac{1}{2}$
				\\
				$0.1$	&	$0.924$	&	$3.076$	&	$\pm 0.538$
				\\
				$0.2$	&	$0.800$	&	$3.200$	&	$\pm 0.600$
				\\
				$0.3$	&	$0.774$	&	$3.226$	&	$\pm 0.613$
				\\
				$0.4$	&	$0.890$	&	$3.115$	&	$\pm 0.557$
				\\
				$0.5$	&	$1.209$	&	$2.791$	&	$\pm 0.395$
			\end{tabular}
		\end{ruledtabular}
	\end{table}
	For very high temperatures one can assume that the system is in the symmetric phase and start the stability analysis from there by succesisvely lowering the temperature until a \gls{sp} to \gls{ip} phase transition or a \gls{sp} to \gls{hbp} phase transition is detected.
	Of course, this is only possible if the phase transition is of second order and the system is in the symmetric phase at high temperatures.

	Independent of the approach, using \cref{eq:bosonic_twopoint_final_T,eq:bosonic_twopoint_final_0} the stability analysis exactly reproduces the phase diagram of the \gls{chgn} model with a \gls{sp}-\gls{ip} phase transition at $T_\mathrm{c} = \frac{\ee^\upgamma}{\uppi}$ for all $\mu$ and $\muchiral$, which is shown in \cref{fig:phasediagram}  (black solid line).
	This is directly clear, if one remembers that the bosonic two-point function of the \gls{gn} model is zero at vanishing chemical potential $\mu$ and vanishing external momentum $q$ exactly for $T_\mathrm{c}$, which actually defines the critical temperature of the phase transition along the $T$-axis.
	Inspecting \cref{eq:bosonic_twopoint_final_T}, where $\mu$ is replaced by $\mu \pm \frac{q}{2}$ \gls{wrt}\ the \gls{gn} two-point function, it is clear that one always finds two combinations of $\mu$ and $q$ for which either the eigenvalues of the Hessian matrix given by \cref{eq:bosonic_twopoint_final_T} vanishes, namely $q = \pm 2 \mu$.
	Hence, we even know the exact momentum of the perturbations, hence the frequency of the modulations of the inhomogeneous condensate at the phase transition.
	It is the wave vector of the chiral spiral, the exact solution.
	Increasing the temperature, both eigenvalues are manifestly positive, which can again be understood from the \gls{gn} model, where the bosonic two-point functions is always positive for $T > T_\mathrm{c}$ at any $\mu$.
	However, for $T < T_\mathrm{c}$ there always exists an interval of non-zero width in $q$ for which the eigenvalues of the two-point function of the \gls{chgn} model are negative.
		\begin{figure}
			\centering
			\includegraphics{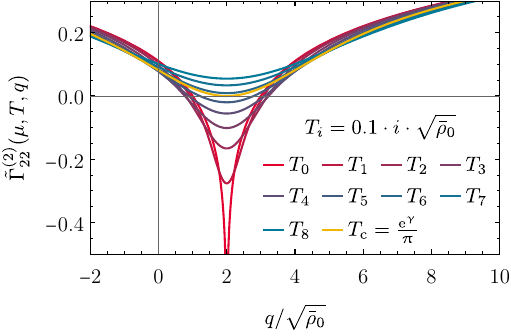}
			\caption{\label{fig:gamma2ofq}%
				Eigenvalue of the bosonic two-point function \labelcref{eq:bosonic_twopoint_final_T} for different $T$ at fixed quark chemical potential $\mu = \sqrt{\bar{\rho}_0}$ as a function of the external momentum $q$.%
			}
		\end{figure}
	The entire situation is captured in \cref{fig:gamma2ofq}, where we show the second eigenvalue of the bosonic two-point function for different $T$ at fixed $\mu = \sqrt{\bar{\rho}_0}$.
	The other eigenvalue as a function of $q$ is obtained by mirroring the plot at $q = 0$.
	For different values of $\mu$ the $q$-profile of the eigenvalues is the same up to a shift in $q$-direction.
	Values of the roots of the second eigenvalue from \cref{fig:gamma2ofq} are provided in \cref{tab:roots_eigenvalue}.

\subsection{Connection to the inner spinodal line of the GN model}

	Inspecting \cref{fig:gamma2ofq} we find that for $T < T_\mathrm{c}$ there always exists an interval in $q$ for which the bosonic two-point functions of the \gls{chgn} model are negative.
	Interestingly, the range in $q$ between the two roots of the eigenvalues of the bosonic two-point function matrix can be directly related to the inner spinodal line of the \gls{gn} model.
	The argument is as follows:
	The spinodal region and second order \gls{sp}-\gls{hbp} transition is defined as the line, where the second derivative of the effective potential vanishes at vanishing field.
	This, however, is exactly the two-point function at vanishing field and vanishing external momentum $q$.
	Since the phase diagram of the \gls{gn} model is actually symmetric \gls{wrt} to the $T$-axis ($\mu \leftrightarrow - \mu$) there are always two values $\pm \mu_\mathrm{spinodal}$ for each $T < T_\mathrm{c}$ where the \gls{gn} two-point function vanishes when evaluated at $q = 0$.
	However, these two values $\pm \mu_\mathrm{spinodal}$ in turn correspond to two combinations of $\mu$ and $q$ for $\pm ( \mu \pm \frac{q}{2} )$ in the \gls{chgn} model, where the \gls{chgn} two-point function vanishes exactly.
	For fixed $\mu$, these roots are the respective end-points of the interval in $q$, within which the bosonic two-point functions of the \gls{chgn} model is negative.
	Hence, the range of momenta that leads to instabilities can be extracted from the inner spinodal line and second order phase transition of the \gls{gn} model.
	Interestingly, in contrast to the stability analysis in the \gls{gn} model in Ref.~\cite{Koenigstein:2021llr}, the minima of the two-point functions are always located at the true wave vector $q$ of the chiral spiral -- the exact solution -- inside the \gls{ip}.
	(To the best of our knowledge these findings go beyond the results of \gls{adpt} in Refs.~\cite{Boehmer:2008uq,Boehmer:2009sw,Thies:2019ejd,Thies:2022tpz}.)

	Of course, everything that is discussed in the results section is valid for arbitrary $\muchiral$, because all expressions are independent of $\muchiral$.


\section{Conclusions and outlook}
\label{sec:conclusions}

	Let us briefly summarize the results of this work and discuss possible future projects.

\subsection{Summary}

	In the introduction we provided a brief motivation of using the stability analysis for the detection of the \gls{ip} in general.
	Despite existing tests and applications of the method, we explained that another test against analytic results is always useful and might reveal new insights into the problem.
	We presented the \gls{chgn} model as a suitable model for this test, because it is a well-known model with a known phase diagram and a known phase transition to an \gls{ip}.
	After recapitulating the basic idea of the stability analysis, we introduced the model as well as the main objects of interest -- the effective potential and the bosonic two-point function.
	Afterwards, we presented the positive results of our work as well as a -- to the best of our knowledge -- unkown interesting connection between the inner spinodal line of the \gls{gn} model and the second order phase transition from the \gls{sp} to the \gls{ip} in the \gls{chgn} model.

\subsection{Conclusions}

	We have successfully applied the stability analysis to the \gls{chgn} model and reproduced the known phase diagram of the model.
	We have shown that for more than one bosonic field the analysis of the eigenvalues of the bosonic two-point function is sufficient to detect the phase transition to an \gls{ip} and the choice of field basis is irrelevant.
	We noticed that diagonalization of the bosonic two-point function can be done completely analytically within the present model, while this is certainly not always the case for models with more involved bosonic field content, see, e.g., Ref.~\cite{Winstel:2024dqu}.
	Hence, in total we presented another test case for the method of the stability analysisand and thereby extended the \gls{adpt}.
	We showed that it is a powerful and reliable tool for the detection of phase transitions even in challenging setups, where little is known about the phase diagram and that the method even provides insights into the shape of the modulations of the inhomogeneous condensates without any ansatz.

\subsection{Outlook}



	It is of course interesting to apply the stability analysis to other models with multiple bosonic fields, combined with arbitrary methods, and in the presence of the bosonic fluctuations.
	This was and is of course already done at several places and in various collaborations.\footnote{However, in some of these works authors did not study the eigenvalues or the determinant of the bosonic two-point function, but single matrix elements, which does not seem to be sufficient according to the present analysis.}
	In particular, for the \gls{chgn} model, the bosonic two-point correlation matrix elements, i.e., the inverse of the quadratic coefficient matrix in \cref{eq:quadratic_coefficient_effective_action}, were studied successfully in Refs.~\cite{Lenz:2021kzo, Lenz:2021vdz,Horie:2021wnn} at finite $N$ on the lattice and another collaboration confirmed their finite-$N$ results with analytic methods \cite{Ciccone:2022zkg, Ciccone:2023pdk}.

	In general, we therefore consider this work as a completion of previous works, especially Refs.~\cite{Sarma:1963,Jacobs:1974ys,Harrington:1974te,Harrington:1974tf,Dashen:1974xz,Schon:2000he,Schon:2000qy,Wolff:1985av,Schnetz:2004vr,Schnetz:2005ih,Schnetz:2005vh,Basar:2009fg,Boehmer:2007ea,Boehmer:2008uq,Boehmer:2009ae,Boehmer:2009sw,Thies:2018qgx,Thies:2019ejd,Koenigstein:2021llr}, and by no means claim that the findings are completely new.
	We still hope that the here presented formulae and discussions is useful for future research.


\begin{acknowledgments}
	A.~K.\ and M.~W.\ thank L.~Pannullo for helpful discussions, comments on the manuscript, and early contributions to the conception of this work.

	A.~K.\ and M.~W.\ thank M.~Thies for helpful comments on the manuscript.
	
	A.~K.\ and M.~W.\ especially thank S.~Floerchinger and M.~Wagner for their general support at the TPI in Jena and the ITP in Frankfurt, respectively.

	M.~W.~ thanks L.~Pannullo for helpful discussions and early contributions to stability analysis computations in related problems. 
	
	M.~W.\ acknowledges support from the \textit{Helmholtz Graduate School for Hadron and Ion Research} and the \textit{Deutsche Forschungsgemeinschaft (DFG, German Research Foundation)} through the Collaborative Research Center TransRegio CRC-TR 211 ``Strong-interaction matter under extreme conditions'' -- project number 315477589 -- TRR 211.

	All plots in this work were produced using \texttt{Mathematica} \cite{Mathematica:13.0} including the \texttt{MaTeX} package \cite{Horvat:matex}.
	The manuscript was prepared with the support of \texttt{GitHubCopilot} \cite{GitHubCopilot2021}.
\end{acknowledgments}


\appendix

\section{Gamma matrices}
\label{app:gamma_matrices}

	For the sake of completeness and to enable the reader to follow the calculations easily, we present the explicit representation of the gamma matrices in Minkowski and Euclidean space.

\subsection{Gamma matrices in Minkowski space}

	In this appendix we present a useful explicit representation of the gamma matrices in Minkowski space.
	The matrices fulfill the anticommutation relation
		\begin{align}
			\big\{ \gamma^\mu, \gamma^\nu \big\}_+ = 2 \eta^{\mu \nu} \mathds{1} \, ,	\label{eq:clifford_algebra_mink}
		\end{align}
	where $\eta^{\mu \nu} = \mathrm{diag} ( 1, - 1 )$.
	We choose the following real representation:
		\begin{align}
			&	\gamma^0 = \sigma_1 =
			\begin{pmatrix}
				0	&	1
				\\
				1	&	0
			\end{pmatrix} \, ,
			&&
			\gamma^1 = \ii \sigma_2 =
			\begin{pmatrix}
				0	&	1
				\\
				- 1	&	0
			\end{pmatrix} \, ,	\label{eq:gamma_matrices_mink}
		\end{align}
	The corresponding chiral gamma matrix is also real and diagonal,
		\begin{align}
			\gammachiral = \gamma^0 \, \gamma^1 =
			\begin{pmatrix}
				- 1	&	0
				\\
				0	&	1
			\end{pmatrix} \, .	\label{eq:gamma_chiral_mink}
		\end{align}
	In addition, we define the chiral projectors
		\begin{align}
			\gammaleft = \, & \tfrac{1}{2} \, ( \mathds{1} - \gammachiral ) =
			\begin{pmatrix}
				1	&	0
				\\
				0	&	0
			\end{pmatrix} \, ,
			\\
			\gammaright = \, & \tfrac{1}{2} \, ( \mathds{1} + \gammachiral ) =
			\begin{pmatrix}
				0	&	0
				\\
				0	&	1
			\end{pmatrix} \, ,	\label{eq:chiral_projectors_mink}
		\end{align}
	which are orthogonal, idempotent, and complete.

\subsection{Gamma matrices in Euclidean space}

	According to our convention for the Wick rotation, the Euclidean gamma matrices obey
		\begin{align}
			\big\{ \gamma^\mu, \gamma^\nu \big\}_+ = 2 \, \delta^{\mu \nu} \, \mathds{1} \, .	\label{eq:clifford_algebra_euclidean}
		\end{align}
	The zeroth gamma matrix is the same, while $\gamma^1 = \ii \gamma^1_\mathrm{M}$, where $\mathrm{M}$ stands for Minkowski space.
		\begin{align}
			&	\gamma^0 =
			\begin{pmatrix}
				0	&	1
				\\
				1	&	0
			\end{pmatrix}
			\, ,
			&&	\gamma^1 =
			\begin{pmatrix}
				0	&	\ii
				\\
				- \ii	&	0
			\end{pmatrix}	\label{eq:gamma_matrices_euclidean}
		\end{align}
	We find that the chiral gamma matrix is the same as in Minkowski space, but its definition in terms of the Euclidean gamma matrices is
		\begin{align}
			\gammachiral = - \ii \gamma^0 \, \gamma^1 =
			\begin{pmatrix}
				- 1	&	0
				\\
				0	&	1
			\end{pmatrix} \, .	\label{eq:gamma_chiral_euclidean}
		\end{align}
	This further implies that also chiral projectors have the same matrix representation in Euclidean and Minkowski space,
		\begin{align}
			&	\gammaleft = \tfrac{1}{2} \, ( \mathds{1} - \gammachiral ) \, ,	&&	\gammaright = \tfrac{1}{2} \, ( \mathds{1} + \gammachiral ) \, ,	\label{eq:chiral_projectors_euclidean}
		\end{align}
	see above.


\section{Evaluation of the traces in the bosonic two-point function}
\label{app:evaluation_of_the_field_theory_traces}

	In this appendix we present the explicit evaluation of the traces in the bosonic two-point function from \cref{eq:unevaluated_bosonic_twopoint}.
	(Throughout this appendix we use that Matsubara sums without explicit limits are understood to be over all Matsubara frequencies.)

	First, we consider the two offdiagonal elements where derivatives of the effective action are taken twice \gls{wrt} $\varphileft$ or $\varphiright$, respectively.
	Inserting the explicit expressions for the bare fermionic propagator \labelcref{eq:fermionic_propagator} and the vertices \labelcref{eq:fermion_boson_vertices}, one finds
		\begin{align}
			& \Tr \Big( ( \S^{(2)} )^{- 1} \S^{(3)}_{\varphileftright ( 0, q_\mathrm{II} )} ( \S^{(2)} )^{- 1} \S^{(3)}_{\varphileftright ( 0, q_\mathrm{I} )} \Big) =	\Vdistance
			\\
			= \, & \tfrac{1}{\beta} \sum_{n_1, \ldots, n_4} \int \frac{\dd p_1 \cdots \dd p_4}{( 2 \uppi )^4} \, \tr \bigg[	\Vdistance	\nonumber
			\\
			& \bigg( \beta \, \delta_{n_1, n_2} \, 2 \uppi \, \delta ( p_1 - p_2 ) \, \frac{\ii ( \nu_{n_2} + \ii \mu ) \, \gamma^0 + \ii ( p_2 + \muchiral ) \, \gamma^1}{( \nu_{n_2} + \ii \mu )^2 + ( p_2 + \muchiral )^2} \bigg)	\Vdistance	\nonumber
			\\
			& \bigg( - \beta \, \delta_{n_2, n_3} \, 2 \uppi \, \delta ( q_\mathrm{II} - p_2 + p_3 ) \, \gammaleftright \bigg)	\Vdistance	\nonumber
			\\
			& \bigg( \beta \, \delta_{n_3, n_4} \, 2 \uppi \, \delta ( p_3 - p_4 ) \, \frac{\ii ( \nu_{n_4} + \ii \mu ) \, \gamma^0 + \ii ( p_4 + \muchiral ) \, \gamma^1}{( \nu_{n_4} + \ii \mu )^2 + ( p_4 + \muchiral )^2} \bigg)	\Vdistance	\nonumber
			\\
			& \bigg( - \beta \, \delta_{n_4, n_1} \, 2 \uppi \, \delta ( q_\mathrm{I} - p_4 + p_1 ) \, \gammaleftright \bigg) \bigg] = 0 \, .	\Vdistance	\nonumber
		\end{align}
	The reason for this is that the Dirac space trace $\tr$ is zero.

	Second, we consider the diagonal terms, which are calculated by taking one derivative of the effective action \gls{wrt} $\varphileft$ and one \gls{wrt} $\varphiright$.
	One finds
		\begin{align}
			&	\Tr \Big( ( \S^{(2)} )^{- 1} \S^{(3)}_{\varphileftright ( 0, q_\mathrm{II} )} ( \S^{(2)} )^{- 1} \S^{(3)}_{\varphirightleft ( 0, q_\mathrm{I} )} \Big) =	\Vdistance
			\\
			= \, & \tfrac{1}{\beta} \sum_{n_1, \ldots, n_4} \int \frac{\dd p_1 \cdots \dd p_4}{( 2 \uppi )^4} \, \tr \bigg[	\Vdistance	\nonumber
			\\
			& \bigg( \beta \, \delta_{n_1, n_2} \, 2 \uppi \, \delta ( p_1 - p_2 ) \, \frac{\ii ( \nu_{n_2} + \ii \mu ) \, \gamma^0 + \ii ( p_2 + \muchiral ) \, \gamma^1}{( \nu_{n_2} + \ii \mu )^2 + ( p_2 + \muchiral )^2} \bigg)	\Vdistance	\nonumber
			\\
			& \bigg( - \beta \, \delta_{n_2, n_3} \, 2 \uppi \, \delta ( q_\mathrm{II} - p_2 + p_3 ) \, \gammaleftright \bigg)	\Vdistance	\nonumber
			\\
			& \bigg( \beta \, \delta_{n_3, n_4} \, 2 \uppi \, \delta ( p_3 - p_4 ) \, \frac{\ii ( \nu_{n_4} + \ii \mu ) \, \gamma^0 + \ii ( p_4 + \muchiral ) \, \gamma^1}{( \nu_{n_4} + \ii \mu )^2 + ( p_4 + \muchiral )^2} \bigg)	\Vdistance	\nonumber
			\\
			& \bigg( - \beta \, \delta_{n_4, n_1} \, 2 \uppi \, \delta ( q_\mathrm{I} - p_4 + p_1 ) \, \gammarightleft \bigg) \bigg] =	\Vdistance	\nonumber
			\\
			= \, & \beta \, 2 \uppi \, \delta ( q_\mathrm{II} + q_\mathrm{I} ) \, \tfrac{1}{\beta} \sum_{n} \int \frac{\dd p}{2 \uppi} \, \tr \bigg[	\Vdistance	\nonumber
			\\
			& \quad \frac{\ii ( \nu_n + \ii \mu ) \, \gamma^0 + \ii ( p + \muchiral ) \, \gamma^1}{( \nu_n + \ii \mu )^2 + ( p + \muchiral )^2} \times	\Vdistance	\nonumber
			\\
			& \quad \times \frac{\ii ( \nu_n + \ii \mu ) \, \gamma^0 + \ii ( p - q_\mathrm{II} + \muchiral ) \, \gamma^1}{( \nu_n + \ii \mu )^2 + ( p - q_\mathrm{II} + \muchiral )^2} \, \gammaleftright \, \bigg] \, .	\Vdistance	\nonumber
		\end{align}
	In the last step, we already moved the chrial projector $\gammaleftright$ to the right.
	Before we further evaluate the Dirac trace, it is convenient to set $q_\mathrm{I} = - q_\mathrm{II} = q$ to ease the notation.
	Furthermore, we shift the spatial momentum integral by the chiral chemical potential $p^\prime = p + \muchiral$ such that the entire expression is $\muchiral$-independent.
	Formally, one should first regularize the momentum integral and perform the shift after the renormalization.
	We have checked that this leads to the same result as long as the system is studied in an infinite volume and the continuum.
	Next, we study the Dirac trace in detail and find,
		\begin{align}
			& \tr \big[ \big( \ii ( \nu_n + \ii \mu ) \, \gamma^0 + \ii p \, \gamma^1 \big) \times	\vdistance
			\\
			& \quad \times \big( \ii ( \nu_n + \ii \mu ) \, \gamma^0 + \ii ( p + q ) \, \gamma^1 \big) \, \gammaleftright \big] =	\vdistance	\nonumber
			\\
			= \, & - \big[ ( \nu_n + \ii \mu )^2 + p \, ( p + q ) \mp \ii q \, ( \nu_n + \ii \mu ) \big] \, .	\vdistance	\nonumber
		\end{align}
	In order to simplify the entire result, one can again shift the momentum integral to $p^\prime = p - \frac{q}{2}$, such that the trace can be rewritten as
			\begin{align}
				\tr ( \ldots ) = \, & - \big[ \big( \nu_n + \ii \big( \mu \mp \tfrac{q}{2} \big) \big)^2 + p^2 \big] \, .	\vdistance	\nonumber
			\end{align}
	Hence, ignoring the overall factor $\beta \, 2 \uppi \, \delta ( q_\mathrm{II} + q_\mathrm{I} )$, the diagonal elements of the bosonic two-point function are
		\begin{align}
			& - \tfrac{1}{\beta} \sum_{n} \int \frac{\dd p}{2 \uppi} \, \big[ \big( \nu_n + \ii \big( \mu \mp \tfrac{q}{2} \big) \big)^2 + p^2 \big] \times	\Vdistance
			\\
			& \times \frac{1}{\big[ ( \nu_n + \ii \mu )^2 + ( p + \frac{q}{2} )^2 \big]\big[ ( \nu_n + \ii \mu )^2 + ( p - \frac{q}{2})^2 \big]}	\Vdistance	\nonumber
			\\
			= \, & - \tfrac{1}{\beta} \sum_{n} \int \frac{\dd p}{2 \uppi} \, \big[ \big( \nu_n + \ii \big( \mu \mp \tfrac{q}{2} \big) \big)^2 + p^2 \big] \times	\Vdistance	\nonumber
			\\
			& \times \frac{1}{\big[ \big( \nu_n + \ii \big( \mu + \tfrac{q}{2} \big) \big)^2 + p^2 \big] \big[ \big( \nu_n + \ii \big( \mu - \tfrac{q}{2} \big) \big)^2 + p^2 \big]} =	\Vdistance	\nonumber
			\\
			= \, & - \tfrac{1}{\beta} \sum_{n} \int \frac{\dd p}{2 \uppi} \, \frac{1}{\big[ \big( \nu_n + \ii \big( \mu \pm \tfrac{q}{2} \big) \big)^2 + p^2 \big]} \, .	\Vdistance	\nonumber
		\end{align}
	Here, however, we notice that the effect of the external momentum is just a shift of the chemical potential and combining our results the diagonal elements -- the eigenvalues of the two-point function are
		\begin{align}
			& \tilde{\Gamma}^{(2)}_{11/22} ( \mu, T, q ) =	\Vdistance	\label{eq:evaluated_bosonic_twopoint}
			\\
			= \, & \tfrac{1}{2 \lambda} - \tfrac{\dimDirac}{2} \, \tfrac{1}{\beta} \sum_{n} \int \frac{\dd p}{2 \uppi} \, \frac{1}{\big[ \big( \nu_n + \ii \big( \mu \pm \tfrac{q}{2} \big) \big)^2 + p^2 \big]}	\Vdistance	\nonumber
		\end{align}
	Here, we have to keep in mind that this expression is only valid in the \gls{sp} or at the phase transition to a \gls{hbp} or \gls{ip}.
	Further evaluation and analysis is presented in the main text.


\FloatBarrier

\bibliography{bib/general.bib,bib/gn.bib,bib/inhomo.bib,bib/instanton.bib,bib/math.bib,bib/numerics.bib,bib/qcd.bib,bib/rg.bib,bib/software.bib,bib/symmetries.bib,bib/thermal_qft.bib,bib/thies.bib,bib/time_crystals.bib,bib/virasoro_algebra.bib,bib/zero-dim-qft.bib}

\begin{thebibliography}{60}%
\makeatletter
\providecommand \@ifxundefined [1]{%
 \@ifx{#1\undefined}
}%
\providecommand \@ifnum [1]{%
 \ifnum #1\expandafter \@firstoftwo
 \else \expandafter \@secondoftwo
 \fi
}%
\providecommand \@ifx [1]{%
 \ifx #1\expandafter \@firstoftwo
 \else \expandafter \@secondoftwo
 \fi
}%
\providecommand \natexlab [1]{#1}%
\providecommand \enquote  [1]{``#1''}%
\providecommand \bibnamefont  [1]{#1}%
\providecommand \bibfnamefont [1]{#1}%
\providecommand \citenamefont [1]{#1}%
\providecommand \href@noop [0]{\@secondoftwo}%
\providecommand \href [0]{\begingroup \@sanitize@url \@href}%
\providecommand \@href[1]{\@@startlink{#1}\@@href}%
\providecommand \@@href[1]{\endgroup#1\@@endlink}%
\providecommand \@sanitize@url [0]{\catcode `\\12\catcode `\$12\catcode
  `\&12\catcode `\#12\catcode `\^12\catcode `\_12\catcode `\%12\relax}%
\providecommand \@@startlink[1]{}%
\providecommand \@@endlink[0]{}%
\providecommand \url  [0]{\begingroup\@sanitize@url \@url }%
\providecommand \@url [1]{\endgroup\@href {#1}{\urlprefix }}%
\providecommand \urlprefix  [0]{URL }%
\providecommand \Eprint [0]{\href }%
\providecommand \doibase [0]{https://doi.org/}%
\providecommand \selectlanguage [0]{\@gobble}%
\providecommand \bibinfo  [0]{\@secondoftwo}%
\providecommand \bibfield  [0]{\@secondoftwo}%
\providecommand \translation [1]{[#1]}%
\providecommand \BibitemOpen [0]{}%
\providecommand \bibitemStop [0]{}%
\providecommand \bibitemNoStop [0]{.\EOS\space}%
\providecommand \EOS [0]{\spacefactor3000\relax}%
\providecommand \BibitemShut  [1]{\csname bibitem#1\endcsname}%
\let\auto@bib@innerbib\@empty
\bibitem [{\citenamefont {Sch\"on}\ and\ \citenamefont
  {Thies}(2000{\natexlab{a}})}]{Schon:2000qy}%
  \BibitemOpen
  \bibfield  {author} {\bibinfo {author} {\bibfnamefont {V.}~\bibnamefont
  {Sch\"on}}\ and\ \bibinfo {author} {\bibfnamefont {M.}~\bibnamefont
  {Thies}},\ }\bibinfo {title} {{2-D model field theories at finite temperature
  and density}},\ in\ \href {https://doi.org/10.1142/9789812810458_0041} {\emph
  {\bibinfo {booktitle} {At The Frontier of Particle Physics: Handbook of QCD,
  Boris Ioffe Festschrift}}},\ Vol.~\bibinfo {volume} {3}\ (\bibinfo
  {publisher} {World Scentific},\ \bibinfo {year} {2000})\ Chap.~\bibinfo
  {chapter} {33}, pp.\ \bibinfo {pages} {1945--2032},\ \Eprint
  {https://arxiv.org/abs/hep-th/0008175} {arXiv:hep-th/0008175} \BibitemShut
  {NoStop}%
\bibitem [{\citenamefont {Boehmer}\ \emph {et~al.}(2008)\citenamefont
  {Boehmer}, \citenamefont {Fritsch}, \citenamefont {Kraus},\ and\
  \citenamefont {Thies}}]{Boehmer:2008uq}%
  \BibitemOpen
  \bibfield  {author} {\bibinfo {author} {\bibfnamefont {C.}~\bibnamefont
  {Boehmer}}, \bibinfo {author} {\bibfnamefont {U.}~\bibnamefont {Fritsch}},
  \bibinfo {author} {\bibfnamefont {S.}~\bibnamefont {Kraus}},\ and\ \bibinfo
  {author} {\bibfnamefont {M.}~\bibnamefont {Thies}},\ }\bibfield  {title}
  {\bibinfo {title} {{Phase structure of the massive chiral Gross-Neveu model
  from Hartree-Fock}},\ }\href {https://doi.org/10.1103/PhysRevD.78.065043}
  {\bibfield  {journal} {\bibinfo  {journal} {Phys. Rev. D}\ }\textbf {\bibinfo
  {volume} {78}},\ \bibinfo {pages} {065043"} (\bibinfo {year} {2008})},\
  \Eprint {https://arxiv.org/abs/0807.2571} {arXiv:0807.2571 [hep-th]}
  \BibitemShut {NoStop}%
\bibitem [{\citenamefont {Boehmer}\ and\ \citenamefont
  {Thies}(2009)}]{Boehmer:2009sw}%
  \BibitemOpen
  \bibfield  {author} {\bibinfo {author} {\bibfnamefont {C.}~\bibnamefont
  {Boehmer}}\ and\ \bibinfo {author} {\bibfnamefont {M.}~\bibnamefont
  {Thies}},\ }\bibfield  {title} {\bibinfo {title} {{Large N solution of
  generalized Gross-Neveu model with two coupling constants}},\ }\href
  {https://doi.org/10.1103/PhysRevD.80.125038} {\bibfield  {journal} {\bibinfo
  {journal} {Phys. Rev. D}\ }\textbf {\bibinfo {volume} {80}},\ \bibinfo
  {pages} {125038} (\bibinfo {year} {2009})},\ \Eprint
  {https://arxiv.org/abs/0909.3714} {arXiv:0909.3714 [hep-th]} \BibitemShut
  {NoStop}%
\bibitem [{\citenamefont {Thies}(2018)}]{Thies:2018qgx}%
  \BibitemOpen
  \bibfield  {author} {\bibinfo {author} {\bibfnamefont {M.}~\bibnamefont
  {Thies}},\ }\bibfield  {title} {\bibinfo {title} {{Chiral spiral in the
  presence of chiral imbalance}},\ }\href
  {https://doi.org/10.1103/PhysRevD.98.096019} {\bibfield  {journal} {\bibinfo
  {journal} {Phys. Rev. D}\ }\textbf {\bibinfo {volume} {98}},\ \bibinfo
  {pages} {096019} (\bibinfo {year} {2018})},\ \Eprint
  {https://arxiv.org/abs/1810.03921} {arXiv:1810.03921 [hep-th]} \BibitemShut
  {NoStop}%
\bibitem [{\citenamefont {Thies}(2022{\natexlab{a}})}]{Thies:2022tpz}%
  \BibitemOpen
  \bibfield  {author} {\bibinfo {author} {\bibfnamefont {M.}~\bibnamefont
  {Thies}},\ }\bibfield  {title} {\bibinfo {title} {{Exact tricritical point
  from next-to-leading-order stability analysis}},\ }\href
  {https://doi.org/10.1103/PhysRevD.105.116003} {\bibfield  {journal} {\bibinfo
   {journal} {Phys. Rev. D}\ }\textbf {\bibinfo {volume} {105}},\ \bibinfo
  {pages} {116003} (\bibinfo {year} {2022}{\natexlab{a}})},\ \Eprint
  {https://arxiv.org/abs/2203.08503} {arXiv:2203.08503 [hep-th]} \BibitemShut
  {NoStop}%
\bibitem [{\citenamefont {Gross}\ and\ \citenamefont
  {Neveu}(1974)}]{Gross:1974jv}%
  \BibitemOpen
  \bibfield  {author} {\bibinfo {author} {\bibfnamefont {D.~J.}\ \bibnamefont
  {Gross}}\ and\ \bibinfo {author} {\bibfnamefont {A.}~\bibnamefont {Neveu}},\
  }\bibfield  {title} {\bibinfo {title} {{Dynamical Symmetry Breaking in
  Asymptotically Free Field Theories}},\ }\href
  {https://doi.org/10.1103/PhysRevD.10.3235} {\bibfield  {journal} {\bibinfo
  {journal} {Phys. Rev. D}\ }\textbf {\bibinfo {volume} {10}},\ \bibinfo
  {pages} {3235} (\bibinfo {year} {1974})}\BibitemShut {NoStop}%
\bibitem [{\citenamefont {Nakano}\ and\ \citenamefont
  {Tatsumi}(2005)}]{Nakano:2004cd}%
  \BibitemOpen
  \bibfield  {author} {\bibinfo {author} {\bibfnamefont {E.}~\bibnamefont
  {Nakano}}\ and\ \bibinfo {author} {\bibfnamefont {T.}~\bibnamefont
  {Tatsumi}},\ }\bibfield  {title} {\bibinfo {title} {{Chiral symmetry and
  density wave in quark matter}},\ }\href
  {https://doi.org/10.1103/PhysRevD.71.114006} {\bibfield  {journal} {\bibinfo
  {journal} {Phys. Rev. D}\ }\textbf {\bibinfo {volume} {71}},\ \bibinfo
  {pages} {114006} (\bibinfo {year} {2005})},\ \Eprint
  {https://arxiv.org/abs/hep-ph/0411350} {arXiv:hep-ph/0411350} \BibitemShut
  {NoStop}%
\bibitem [{\citenamefont {Abuki}\ \emph {et~al.}(2012)\citenamefont {Abuki},
  \citenamefont {Ishibashi},\ and\ \citenamefont {Suzuki}}]{Abuki:2011pf}%
  \BibitemOpen
  \bibfield  {author} {\bibinfo {author} {\bibfnamefont {H.}~\bibnamefont
  {Abuki}}, \bibinfo {author} {\bibfnamefont {D.}~\bibnamefont {Ishibashi}},\
  and\ \bibinfo {author} {\bibfnamefont {K.}~\bibnamefont {Suzuki}},\
  }\bibfield  {title} {\bibinfo {title} {{Crystalline chiral condensates off
  the tricritical point in a generalized Ginzburg-Landau approach}},\ }\href
  {https://doi.org/10.1103/PhysRevD.85.074002} {\bibfield  {journal} {\bibinfo
  {journal} {Phys. Rev. D}\ }\textbf {\bibinfo {volume} {85}},\ \bibinfo
  {pages} {074002} (\bibinfo {year} {2012})},\ \Eprint
  {https://arxiv.org/abs/1109.1615} {arXiv:1109.1615 [hep-ph]} \BibitemShut
  {NoStop}%
\bibitem [{\citenamefont {de~Forcrand}\ and\ \citenamefont
  {Wenger}(2006)}]{deForcrand:2006zz}%
  \BibitemOpen
  \bibfield  {author} {\bibinfo {author} {\bibfnamefont {P.}~\bibnamefont
  {de~Forcrand}}\ and\ \bibinfo {author} {\bibfnamefont {U.}~\bibnamefont
  {Wenger}},\ }\bibfield  {title} {\bibinfo {title} {{New baryon matter in the
  lattice Gross-Neveu model}},\ }\href {https://doi.org/10.22323/1.032.0152}
  {\bibfield  {journal} {\bibinfo  {journal} {PoS}\ }\textbf {\bibinfo {volume}
  {LAT2006}},\ \bibinfo {pages} {152} (\bibinfo {year} {2006})},\ \Eprint
  {https://arxiv.org/abs/hep-lat/0610117} {arXiv:hep-lat/0610117} \BibitemShut
  {NoStop}%
\bibitem [{\citenamefont {Tripolt}\ \emph {et~al.}(2018)\citenamefont
  {Tripolt}, \citenamefont {Schaefer}, \citenamefont {von Smekal},\ and\
  \citenamefont {Wambach}}]{Tripolt:2017zgc}%
  \BibitemOpen
  \bibfield  {author} {\bibinfo {author} {\bibfnamefont {R.-A.}\ \bibnamefont
  {Tripolt}}, \bibinfo {author} {\bibfnamefont {B.-J.}\ \bibnamefont
  {Schaefer}}, \bibinfo {author} {\bibfnamefont {L.}~\bibnamefont {von
  Smekal}},\ and\ \bibinfo {author} {\bibfnamefont {J.}~\bibnamefont
  {Wambach}},\ }\bibfield  {title} {\bibinfo {title} {{Low-temperature behavior
  of the quark-meson model}},\ }\href
  {https://doi.org/10.1103/PhysRevD.97.034022} {\bibfield  {journal} {\bibinfo
  {journal} {Phys. Rev. D}\ }\textbf {\bibinfo {volume} {97}},\ \bibinfo
  {pages} {034022} (\bibinfo {year} {2018})},\ \Eprint
  {https://arxiv.org/abs/1709.05991} {arXiv:1709.05991 [hep-ph]} \BibitemShut
  {NoStop}%
\bibitem [{\citenamefont {Buballa}\ \emph {et~al.}(2021)\citenamefont
  {Buballa}, \citenamefont {Kurth}, \citenamefont {Wagner},\ and\ \citenamefont
  {Winstel}}]{Buballa:2020nsi}%
  \BibitemOpen
  \bibfield  {author} {\bibinfo {author} {\bibfnamefont {M.}~\bibnamefont
  {Buballa}}, \bibinfo {author} {\bibfnamefont {L.}~\bibnamefont {Kurth}},
  \bibinfo {author} {\bibfnamefont {M.}~\bibnamefont {Wagner}},\ and\ \bibinfo
  {author} {\bibfnamefont {M.}~\bibnamefont {Winstel}},\ }\bibfield  {title}
  {\bibinfo {title} {{Regulator dependence of inhomogeneous phases in the ( 2+1
  )-dimensional Gross-Neveu model}},\ }\href
  {https://doi.org/10.1103/PhysRevD.103.034503} {\bibfield  {journal} {\bibinfo
   {journal} {Phys. Rev. D}\ }\textbf {\bibinfo {volume} {103}},\ \bibinfo
  {pages} {034503} (\bibinfo {year} {2021})},\ \Eprint
  {https://arxiv.org/abs/2012.09588} {arXiv:2012.09588 [hep-lat]} \BibitemShut
  {NoStop}%
\bibitem [{\citenamefont {Buballa}\ \emph {et~al.}(2020)\citenamefont
  {Buballa}, \citenamefont {Carignano},\ and\ \citenamefont
  {Kurth}}]{Buballa:2020xaa}%
  \BibitemOpen
  \bibfield  {author} {\bibinfo {author} {\bibfnamefont {M.}~\bibnamefont
  {Buballa}}, \bibinfo {author} {\bibfnamefont {S.}~\bibnamefont {Carignano}},\
  and\ \bibinfo {author} {\bibfnamefont {L.}~\bibnamefont {Kurth}},\ }\bibfield
   {title} {\bibinfo {title} {{Inhomogeneous phases in the quark-meson model
  with explicit chiral-symmetry breaking}},\ }\href
  {https://doi.org/10.1140/epjst/e2020-000101-x"} {\bibfield  {journal}
  {\bibinfo  {journal} {Eur. Phys. J. ST}\ }\textbf {\bibinfo {volume} {229}},\
  \bibinfo {pages} {3371} (\bibinfo {year} {2020})},\ \Eprint
  {https://arxiv.org/abs/2006.02133} {arXiv:2006.02133 [hep-ph]} \BibitemShut
  {NoStop}%
\bibitem [{\citenamefont {Pannullo}\ \emph {et~al.}(2022)\citenamefont
  {Pannullo}, \citenamefont {Wagner},\ and\ \citenamefont
  {Winstel}}]{Pannullo:2021edr}%
  \BibitemOpen
  \bibfield  {author} {\bibinfo {author} {\bibfnamefont {L.}~\bibnamefont
  {Pannullo}}, \bibinfo {author} {\bibfnamefont {M.}~\bibnamefont {Wagner}},\
  and\ \bibinfo {author} {\bibfnamefont {M.}~\bibnamefont {Winstel}},\
  }\bibfield  {title} {\bibinfo {title} {{Inhomogeneous Phases in the Chirally
  Imbalanced 2 + 1-Dimensional Gross-Neveu Model and Their Absence in the
  Continuum Limit}},\ }\href {https://doi.org/10.3390/sym14020265} {\bibfield
  {journal} {\bibinfo  {journal} {Symmetry}\ }\textbf {\bibinfo {volume}
  {14}},\ \bibinfo {pages} {265} (\bibinfo {year} {2022})},\ \Eprint
  {https://arxiv.org/abs/2112.11183} {arXiv:2112.11183 [hep-lat]} \BibitemShut
  {NoStop}%
\bibitem [{\citenamefont {Pannullo}\ \emph {et~al.}(2023)\citenamefont
  {Pannullo}, \citenamefont {Wagner},\ and\ \citenamefont
  {Winstel}}]{Pannullo:2022eqh}%
  \BibitemOpen
  \bibfield  {author} {\bibinfo {author} {\bibfnamefont {L.}~\bibnamefont
  {Pannullo}}, \bibinfo {author} {\bibfnamefont {M.}~\bibnamefont {Wagner}},\
  and\ \bibinfo {author} {\bibfnamefont {M.}~\bibnamefont {Winstel}},\
  }\bibfield  {title} {\bibinfo {title} {{Inhomogeneous phases in the
  3+1-dimensional Nambu-Jona-Lasinio model and their dependence on the
  regularization scheme}},\ }\href {https://doi.org/10.22323/1.430.0156}
  {\bibfield  {journal} {\bibinfo  {journal} {PoS}\ }\textbf {\bibinfo {volume}
  {LATTICE2022}},\ \bibinfo {pages} {156} (\bibinfo {year} {2023})},\ \Eprint
  {https://arxiv.org/abs/2212.05783} {arXiv:2212.05783 [hep-lat]} \BibitemShut
  {NoStop}%
\bibitem [{\citenamefont {Pannullo}\ and\ \citenamefont
  {Winstel}(2023)}]{Pannullo:2023one}%
  \BibitemOpen
  \bibfield  {author} {\bibinfo {author} {\bibfnamefont {L.}~\bibnamefont
  {Pannullo}}\ and\ \bibinfo {author} {\bibfnamefont {M.}~\bibnamefont
  {Winstel}},\ }\bibfield  {title} {\bibinfo {title} {{Absence of inhomogeneous
  chiral phases in (2+1)-dimensional four-fermion and Yukawa models}},\ }\href
  {https://doi.org/10.1103/PhysRevD.108.036011} {\bibfield  {journal} {\bibinfo
   {journal} {Phys. Rev. D}\ }\textbf {\bibinfo {volume} {108}},\ \bibinfo
  {pages} {036011} (\bibinfo {year} {2023})},\ \Eprint
  {https://arxiv.org/abs/2305.09444} {arXiv:2305.09444 [hep-ph]} \BibitemShut
  {NoStop}%
\bibitem [{\citenamefont {Pannullo}(2023)}]{Pannullo:2023cat}%
  \BibitemOpen
  \bibfield  {author} {\bibinfo {author} {\bibfnamefont {L.}~\bibnamefont
  {Pannullo}},\ }\bibfield  {title} {\bibinfo {title} {{Inhomogeneous
  condensation in the Gross-Neveu model in noninteger spatial dimensions
  1\ensuremath{\leq}d\ensuremath{<}3}},\ }\href
  {https://doi.org/10.1103/PhysRevD.108.036022} {\bibfield  {journal} {\bibinfo
   {journal} {Phys. Rev. D}\ }\textbf {\bibinfo {volume} {108}},\ \bibinfo
  {pages} {036022} (\bibinfo {year} {2023})},\ \Eprint
  {https://arxiv.org/abs/2306.16290} {arXiv:2306.16290 [hep-ph]} \BibitemShut
  {NoStop}%
\bibitem [{\citenamefont {Koenigstein}\ and\ \citenamefont
  {Pannullo}(2024)}]{Koenigstein:2023yzv}%
  \BibitemOpen
  \bibfield  {author} {\bibinfo {author} {\bibfnamefont {A.}~\bibnamefont
  {Koenigstein}}\ and\ \bibinfo {author} {\bibfnamefont {L.}~\bibnamefont
  {Pannullo}},\ }\bibfield  {title} {\bibinfo {title} {{Inhomogeneous
  condensation in the Gross-Neveu model in noninteger spatial dimensions
  1\ensuremath{\leq}d\ensuremath{<}3. II. Nonzero temperature and chemical
  potential}},\ }\href {https://doi.org/10.1103/PhysRevD.109.056015} {\bibfield
   {journal} {\bibinfo  {journal} {Phys. Rev. D}\ }\textbf {\bibinfo {volume}
  {109}},\ \bibinfo {pages} {056015} (\bibinfo {year} {2024})},\ \Eprint
  {https://arxiv.org/abs/2312.04904} {arXiv:2312.04904 [hep-ph]} \BibitemShut
  {NoStop}%
\bibitem [{\citenamefont {Winstel}(2024)}]{Winstel:2024dqu}%
  \BibitemOpen
  \bibfield  {author} {\bibinfo {author} {\bibfnamefont {M.}~\bibnamefont
  {Winstel}},\ }\href@noop {} {\bibinfo {title} {{Spatially oscillating
  correlation functions in $\left(2+1\right)$-dimensional four-fermion models:
  The mixing of scalar and vector modes at finite density}}} (\bibinfo {year}
  {2024}),\ \Eprint {https://arxiv.org/abs/2403.07430} {arXiv:2403.07430
  [hep-ph]} \BibitemShut {NoStop}%
\bibitem [{\citenamefont {Braun}\ \emph {et~al.}(2015)\citenamefont {Braun},
  \citenamefont {Finkbeiner}, \citenamefont {Karbstein},\ and\ \citenamefont
  {Roscher}}]{Braun:2014fga}%
  \BibitemOpen
  \bibfield  {author} {\bibinfo {author} {\bibfnamefont {J.}~\bibnamefont
  {Braun}}, \bibinfo {author} {\bibfnamefont {S.}~\bibnamefont {Finkbeiner}},
  \bibinfo {author} {\bibfnamefont {F.}~\bibnamefont {Karbstein}},\ and\
  \bibinfo {author} {\bibfnamefont {D.}~\bibnamefont {Roscher}},\ }\bibfield
  {title} {\bibinfo {title} {{Search for inhomogeneous phases in fermionic
  models}},\ }\href {https://doi.org/10.1103/PhysRevD.91.116006} {\bibfield
  {journal} {\bibinfo  {journal} {Phys. Rev. D}\ }\textbf {\bibinfo {volume}
  {91}},\ \bibinfo {pages} {116006} (\bibinfo {year} {2015})},\ \Eprint
  {https://arxiv.org/abs/1410.8181} {arXiv:1410.8181 [hep-ph]} \BibitemShut
  {NoStop}%
\bibitem [{\citenamefont {Koenigstein}\ \emph {et~al.}(2022)\citenamefont
  {Koenigstein}, \citenamefont {Pannullo}, \citenamefont {Rechenberger},
  \citenamefont {Steil},\ and\ \citenamefont {Winstel}}]{Koenigstein:2021llr}%
  \BibitemOpen
  \bibfield  {author} {\bibinfo {author} {\bibfnamefont {A.}~\bibnamefont
  {Koenigstein}}, \bibinfo {author} {\bibfnamefont {L.}~\bibnamefont
  {Pannullo}}, \bibinfo {author} {\bibfnamefont {S.}~\bibnamefont
  {Rechenberger}}, \bibinfo {author} {\bibfnamefont {M.~J.}\ \bibnamefont
  {Steil}},\ and\ \bibinfo {author} {\bibfnamefont {M.}~\bibnamefont
  {Winstel}},\ }\bibfield  {title} {\bibinfo {title} {{Detecting inhomogeneous
  chiral condensation from the bosonic two-point function in the (1 +
  1)-dimensional Gross\textendash{}Neveu model in the mean-field
  approximation*}},\ }\href {https://doi.org/10.1088/1751-8121/ac820a}
  {\bibfield  {journal} {\bibinfo  {journal} {J. Phys. A}\ }\textbf {\bibinfo
  {volume} {55}},\ \bibinfo {pages} {375402} (\bibinfo {year} {2022})},\
  \Eprint {https://arxiv.org/abs/2112.07024} {arXiv:2112.07024 [hep-ph]}
  \BibitemShut {NoStop}%
\bibitem [{\citenamefont {Buballa}\ and\ \citenamefont
  {Carignano}(2015)}]{Buballa:2014tba}%
  \BibitemOpen
  \bibfield  {author} {\bibinfo {author} {\bibfnamefont {M.}~\bibnamefont
  {Buballa}}\ and\ \bibinfo {author} {\bibfnamefont {S.}~\bibnamefont
  {Carignano}},\ }\bibfield  {title} {\bibinfo {title} {{Inhomogeneous chiral
  condensates}},\ }\href {https://doi.org/10.1016/j.ppnp.2014.11.001}
  {\bibfield  {journal} {\bibinfo  {journal} {Prog. Part. Nucl. Phys.}\
  }\textbf {\bibinfo {volume} {81}},\ \bibinfo {pages} {39} (\bibinfo {year}
  {2015})},\ \Eprint {https://arxiv.org/abs/1406.1367} {arXiv:1406.1367
  [hep-ph]} \BibitemShut {NoStop}%
\bibitem [{\citenamefont {Thies}(2020)}]{Thies:2019ejd}%
  \BibitemOpen
  \bibfield  {author} {\bibinfo {author} {\bibfnamefont {M.}~\bibnamefont
  {Thies}},\ }\bibfield  {title} {\bibinfo {title} {{Phase structure of the $(1
  + 1)$-dimensional Nambu-Jona-Lasinio model with isospin}},\ }\href
  {https://doi.org/10.1103/PhysRevD.101.014010} {\bibfield  {journal} {\bibinfo
   {journal} {Phys. Rev. D}\ }\textbf {\bibinfo {volume} {101}},\ \bibinfo
  {pages} {014010} (\bibinfo {year} {2020})},\ \Eprint
  {https://arxiv.org/abs/1911.11439} {arXiv:1911.11439 [hep-th]} \BibitemShut
  {NoStop}%
\bibitem [{\citenamefont {Sarma}(1963)}]{Sarma:1963}%
  \BibitemOpen
  \bibfield  {author} {\bibinfo {author} {\bibfnamefont {G.}~\bibnamefont
  {Sarma}},\ }\bibfield  {title} {\bibinfo {title} {On the influence of a
  uniform exchange field acting on the spins of the conduction electrons in a
  superconductor},\ }\href {https://doi.org/10.1016/0022-3697(63)90007-6}
  {\bibfield  {journal} {\bibinfo  {journal} {Journal of Physics and Chemistry
  of Solids}\ }\textbf {\bibinfo {volume} {24}},\ \bibinfo {pages} {1029}
  (\bibinfo {year} {1963})}\BibitemShut {NoStop}%
\bibitem [{\citenamefont {Jacobs}(1974)}]{Jacobs:1974ys}%
  \BibitemOpen
  \bibfield  {author} {\bibinfo {author} {\bibfnamefont {L.}~\bibnamefont
  {Jacobs}},\ }\bibfield  {title} {\bibinfo {title} {{Critical behavior in a
  class of $O(N)$-invariant field theories in two dimensions}},\ }\href
  {https://doi.org/10.1103/PhysRevD.10.3956} {\bibfield  {journal} {\bibinfo
  {journal} {Phys. Rev. D}\ }\textbf {\bibinfo {volume} {10}},\ \bibinfo
  {pages} {3956} (\bibinfo {year} {1974})}\BibitemShut {NoStop}%
\bibitem [{\citenamefont {Harrington}\ and\ \citenamefont
  {Yildiz}(1975{\natexlab{a}})}]{Harrington:1974te}%
  \BibitemOpen
  \bibfield  {author} {\bibinfo {author} {\bibfnamefont {B.~J.}\ \bibnamefont
  {Harrington}}\ and\ \bibinfo {author} {\bibfnamefont {A.}~\bibnamefont
  {Yildiz}},\ }\bibfield  {title} {\bibinfo {title} {{Chiral Symmetry Behavior
  at Large Densities}},\ }\href {https://doi.org/10.1103/PhysRevD.11.1705}
  {\bibfield  {journal} {\bibinfo  {journal} {Phys. Rev. D}\ }\textbf {\bibinfo
  {volume} {11}},\ \bibinfo {pages} {1705} (\bibinfo {year}
  {1975}{\natexlab{a}})}\BibitemShut {NoStop}%
\bibitem [{\citenamefont {Harrington}\ and\ \citenamefont
  {Yildiz}(1975{\natexlab{b}})}]{Harrington:1974tf}%
  \BibitemOpen
  \bibfield  {author} {\bibinfo {author} {\bibfnamefont {B.~J.}\ \bibnamefont
  {Harrington}}\ and\ \bibinfo {author} {\bibfnamefont {A.}~\bibnamefont
  {Yildiz}},\ }\bibfield  {title} {\bibinfo {title} {{Restoration of
  Dynamically Broken Symmetries at Finite Temperature}},\ }\href
  {https://doi.org/10.1103/PhysRevD.11.779} {\bibfield  {journal} {\bibinfo
  {journal} {Phys. Rev. D}\ }\textbf {\bibinfo {volume} {11}},\ \bibinfo
  {pages} {779} (\bibinfo {year} {1975}{\natexlab{b}})}\BibitemShut {NoStop}%
\bibitem [{\citenamefont {Dashen}\ \emph {et~al.}(1975)\citenamefont {Dashen},
  \citenamefont {Ma},\ and\ \citenamefont {Rajaraman}}]{Dashen:1974xz}%
  \BibitemOpen
  \bibfield  {author} {\bibinfo {author} {\bibfnamefont {R.~F.}\ \bibnamefont
  {Dashen}}, \bibinfo {author} {\bibfnamefont {S.-k.}\ \bibnamefont {Ma}},\
  and\ \bibinfo {author} {\bibfnamefont {R.}~\bibnamefont {Rajaraman}},\
  }\bibfield  {title} {\bibinfo {title} {{Finite temperature behavior of a
  relativistic field theory with dynamical symmetry breaking}},\ }\href
  {https://doi.org/10.1103/PhysRevD.11.1499} {\bibfield  {journal} {\bibinfo
  {journal} {Phys. Rev. D}\ }\textbf {\bibinfo {volume} {11}},\ \bibinfo
  {pages} {1499} (\bibinfo {year} {1975})}\BibitemShut {NoStop}%
\bibitem [{\citenamefont {Sch\"on}\ and\ \citenamefont
  {Thies}(2000{\natexlab{b}})}]{Schon:2000he}%
  \BibitemOpen
  \bibfield  {author} {\bibinfo {author} {\bibfnamefont {V.}~\bibnamefont
  {Sch\"on}}\ and\ \bibinfo {author} {\bibfnamefont {M.}~\bibnamefont
  {Thies}},\ }\bibfield  {title} {\bibinfo {title} {{Emergence of Skyrme
  crystal in Gross-Neveu and 't Hooft models at finite density}},\ }\href
  {https://doi.org/10.1103/PhysRevD.62.096002} {\bibfield  {journal} {\bibinfo
  {journal} {Phys. Rev. D}\ }\textbf {\bibinfo {volume} {62}},\ \bibinfo
  {pages} {096002} (\bibinfo {year} {2000}{\natexlab{b}})},\ \Eprint
  {https://arxiv.org/abs/hep-th/0003195} {arXiv:hep-th/0003195} \BibitemShut
  {NoStop}%
\bibitem [{\citenamefont {Wolff}(1985)}]{Wolff:1985av}%
  \BibitemOpen
  \bibfield  {author} {\bibinfo {author} {\bibfnamefont {U.}~\bibnamefont
  {Wolff}},\ }\bibfield  {title} {\bibinfo {title} {{The phase diagram of the
  infinite-N Gross-Neveu model at finite temperature and chemical potential}},\
  }\href {https://doi.org/10.1016/0370-2693(85)90671-9} {\bibfield  {journal}
  {\bibinfo  {journal} {Phys. Lett. B}\ }\textbf {\bibinfo {volume} {157}},\
  \bibinfo {pages} {303} (\bibinfo {year} {1985})}\BibitemShut {NoStop}%
\bibitem [{\citenamefont {Schnetz}\ \emph {et~al.}(2004)\citenamefont
  {Schnetz}, \citenamefont {Thies},\ and\ \citenamefont
  {Urlichs}}]{Schnetz:2004vr}%
  \BibitemOpen
  \bibfield  {author} {\bibinfo {author} {\bibfnamefont {O.}~\bibnamefont
  {Schnetz}}, \bibinfo {author} {\bibfnamefont {M.}~\bibnamefont {Thies}},\
  and\ \bibinfo {author} {\bibfnamefont {K.}~\bibnamefont {Urlichs}},\
  }\bibfield  {title} {\bibinfo {title} {{Phase diagram of the Gross-Neveu
  model: Exact results and condensed matter precursors}},\ }\href
  {https://doi.org/10.1016/j.aop.2004.06.009} {\bibfield  {journal} {\bibinfo
  {journal} {Annals Phys.}\ }\textbf {\bibinfo {volume} {314}},\ \bibinfo
  {pages} {425} (\bibinfo {year} {2004})},\ \Eprint
  {https://arxiv.org/abs/hep-th/0402014} {arXiv:hep-th/0402014} \BibitemShut
  {NoStop}%
\bibitem [{\citenamefont {Schnetz}\ \emph {et~al.}(2006)\citenamefont
  {Schnetz}, \citenamefont {Thies},\ and\ \citenamefont
  {Urlichs}}]{Schnetz:2005ih}%
  \BibitemOpen
  \bibfield  {author} {\bibinfo {author} {\bibfnamefont {O.}~\bibnamefont
  {Schnetz}}, \bibinfo {author} {\bibfnamefont {M.}~\bibnamefont {Thies}},\
  and\ \bibinfo {author} {\bibfnamefont {K.}~\bibnamefont {Urlichs}},\
  }\bibfield  {title} {\bibinfo {title} {{Full phase diagram of the massive
  Gross-Neveu model}},\ }\href {https://doi.org/10.1016/j.aop.2005.12.007}
  {\bibfield  {journal} {\bibinfo  {journal} {Annals Phys.}\ }\textbf {\bibinfo
  {volume} {321}},\ \bibinfo {pages} {2604} (\bibinfo {year} {2006})},\ \Eprint
  {https://arxiv.org/abs/hep-th/0511206} {arXiv:hep-th/0511206} \BibitemShut
  {NoStop}%
\bibitem [{\citenamefont {Schnetz}\ \emph {et~al.}(2005)\citenamefont
  {Schnetz}, \citenamefont {Thies},\ and\ \citenamefont
  {Urlichs}}]{Schnetz:2005vh}%
  \BibitemOpen
  \bibfield  {author} {\bibinfo {author} {\bibfnamefont {O.}~\bibnamefont
  {Schnetz}}, \bibinfo {author} {\bibfnamefont {M.}~\bibnamefont {Thies}},\
  and\ \bibinfo {author} {\bibfnamefont {K.}~\bibnamefont {Urlichs}},\
  }\href@noop {} {\bibinfo {title} {{The Phase diagram of the massive
  Gross-Neveu model, revisited}}} (\bibinfo {year} {2005}),\ \Eprint
  {https://arxiv.org/abs/hep-th/0507120} {arXiv:hep-th/0507120} \BibitemShut
  {NoStop}%
\bibitem [{\citenamefont {Basar}\ \emph {et~al.}(2009)\citenamefont {Basar},
  \citenamefont {Dunne},\ and\ \citenamefont {Thies}}]{Basar:2009fg}%
  \BibitemOpen
  \bibfield  {author} {\bibinfo {author} {\bibfnamefont {G.}~\bibnamefont
  {Basar}}, \bibinfo {author} {\bibfnamefont {G.~V.}\ \bibnamefont {Dunne}},\
  and\ \bibinfo {author} {\bibfnamefont {M.}~\bibnamefont {Thies}},\ }\bibfield
   {title} {\bibinfo {title} {{Inhomogeneous condensates in the thermodynamics
  of the chiral ${\mathrm{NJL}}_{2}$ model}},\ }\href
  {https://doi.org/10.1103/PhysRevD.79.105012} {\bibfield  {journal} {\bibinfo
  {journal} {Phys. Rev. D}\ }\textbf {\bibinfo {volume} {79}},\ \bibinfo
  {pages} {105012} (\bibinfo {year} {2009})},\ \Eprint
  {https://arxiv.org/abs/0903.1868} {arXiv:0903.1868 [hep-th]} \BibitemShut
  {NoStop}%
\bibitem [{\citenamefont {Boehmer}\ \emph {et~al.}(2007)\citenamefont
  {Boehmer}, \citenamefont {Thies},\ and\ \citenamefont
  {Urlichs}}]{Boehmer:2007ea}%
  \BibitemOpen
  \bibfield  {author} {\bibinfo {author} {\bibfnamefont {C.}~\bibnamefont
  {Boehmer}}, \bibinfo {author} {\bibfnamefont {M.}~\bibnamefont {Thies}},\
  and\ \bibinfo {author} {\bibfnamefont {K.}~\bibnamefont {Urlichs}},\
  }\bibfield  {title} {\bibinfo {title} {{Tricritical behavior of the massive
  chiral Gross-Neveu model}},\ }\href
  {https://doi.org/10.1103/PhysRevD.75.105017} {\bibfield  {journal} {\bibinfo
  {journal} {Phys. Rev. D}\ }\textbf {\bibinfo {volume} {75}},\ \bibinfo
  {pages} {105017} (\bibinfo {year} {2007})},\ \Eprint
  {https://arxiv.org/abs/hep-th/0702201} {arXiv:hep-th/0702201} \BibitemShut
  {NoStop}%
\bibitem [{\citenamefont {Boehmer}\ and\ \citenamefont
  {Thies}(2010)}]{Boehmer:2009ae}%
  \BibitemOpen
  \bibfield  {author} {\bibinfo {author} {\bibfnamefont {C.}~\bibnamefont
  {Boehmer}}\ and\ \bibinfo {author} {\bibfnamefont {M.}~\bibnamefont
  {Thies}},\ }\bibfield  {title} {\bibinfo {title} {{Competing mechanisms of
  chiral symmetry breaking in a generalized Gross-Neveu model}},\ }\href
  {https://doi.org/10.1103/PhysRevD.81.105027} {\bibfield  {journal} {\bibinfo
  {journal} {Phys. Rev. D}\ }\textbf {\bibinfo {volume} {81}},\ \bibinfo
  {pages} {105027} (\bibinfo {year} {2010})},\ \Eprint
  {https://arxiv.org/abs/0912.2664} {arXiv:0912.2664 [hep-th]} \BibitemShut
  {NoStop}%
\bibitem [{\citenamefont {Ciccone}\ \emph {et~al.}(2024)\citenamefont
  {Ciccone}, \citenamefont {Di~Pietro},\ and\ \citenamefont
  {Serone}}]{Ciccone:2023pdk}%
  \BibitemOpen
  \bibfield  {author} {\bibinfo {author} {\bibfnamefont {R.}~\bibnamefont
  {Ciccone}}, \bibinfo {author} {\bibfnamefont {L.}~\bibnamefont {Di~Pietro}},\
  and\ \bibinfo {author} {\bibfnamefont {M.}~\bibnamefont {Serone}},\
  }\bibfield  {title} {\bibinfo {title} {{Anomalies and persistent order in the
  chiral Gross-Neveu model}},\ }\href {https://doi.org/10.1007/JHEP02(2024)211}
  {\bibfield  {journal} {\bibinfo  {journal} {JHEP}\ }\textbf {\bibinfo
  {volume} {02}},\ \bibinfo {pages} {211}},\ \Eprint
  {https://arxiv.org/abs/2312.13756} {arXiv:2312.13756 [hep-th]} \BibitemShut
  {NoStop}%
\bibitem [{\citenamefont {Lenz}\ \emph
  {et~al.}(2020{\natexlab{a}})\citenamefont {Lenz}, \citenamefont {Pannullo},
  \citenamefont {Wagner}, \citenamefont {Wellegehausen},\ and\ \citenamefont
  {Wipf}}]{Lenz:2020bxk}%
  \BibitemOpen
  \bibfield  {author} {\bibinfo {author} {\bibfnamefont {J.}~\bibnamefont
  {Lenz}}, \bibinfo {author} {\bibfnamefont {L.}~\bibnamefont {Pannullo}},
  \bibinfo {author} {\bibfnamefont {M.}~\bibnamefont {Wagner}}, \bibinfo
  {author} {\bibfnamefont {B.}~\bibnamefont {Wellegehausen}},\ and\ \bibinfo
  {author} {\bibfnamefont {A.}~\bibnamefont {Wipf}},\ }\bibfield  {title}
  {\bibinfo {title} {{Inhomogeneous phases in the Gross-Neveu model in 1+1
  dimensions at finite number of flavors}},\ }\href
  {https://doi.org/10.1103/PhysRevD.101.094512} {\bibfield  {journal} {\bibinfo
   {journal} {Phys. Rev. D}\ }\textbf {\bibinfo {volume} {101}},\ \bibinfo
  {pages} {094512} (\bibinfo {year} {2020}{\natexlab{a}})},\ \Eprint
  {https://arxiv.org/abs/2004.00295} {arXiv:2004.00295 [hep-lat]} \BibitemShut
  {NoStop}%
\bibitem [{\citenamefont {Lenz}\ \emph
  {et~al.}(2020{\natexlab{b}})\citenamefont {Lenz}, \citenamefont {Pannullo},
  \citenamefont {Wagner}, \citenamefont {Wellegehausen},\ and\ \citenamefont
  {Wipf}}]{Lenz:2020cuv}%
  \BibitemOpen
  \bibfield  {author} {\bibinfo {author} {\bibfnamefont {J.~J.}\ \bibnamefont
  {Lenz}}, \bibinfo {author} {\bibfnamefont {L.}~\bibnamefont {Pannullo}},
  \bibinfo {author} {\bibfnamefont {M.}~\bibnamefont {Wagner}}, \bibinfo
  {author} {\bibfnamefont {B.~H.}\ \bibnamefont {Wellegehausen}},\ and\
  \bibinfo {author} {\bibfnamefont {A.}~\bibnamefont {Wipf}},\ }\bibfield
  {title} {\bibinfo {title} {{Baryons in the Gross-Neveu model in 1+1
  dimensions at finite number of flavors}},\ }\href
  {https://doi.org/10.1103/PhysRevD.102.114501} {\bibfield  {journal} {\bibinfo
   {journal} {Phys. Rev. D}\ }\textbf {\bibinfo {volume} {102}},\ \bibinfo
  {pages} {114501} (\bibinfo {year} {2020}{\natexlab{b}})},\ \Eprint
  {https://arxiv.org/abs/2007.08382} {arXiv:2007.08382 [hep-lat]} \BibitemShut
  {NoStop}%
\bibitem [{\citenamefont {Stoll}\ \emph {et~al.}(2021)\citenamefont {Stoll},
  \citenamefont {Zorbach}, \citenamefont {Koenigstein}, \citenamefont {Steil},\
  and\ \citenamefont {Rechenberger}}]{Stoll:2021ori}%
  \BibitemOpen
  \bibfield  {author} {\bibinfo {author} {\bibfnamefont {J.}~\bibnamefont
  {Stoll}}, \bibinfo {author} {\bibfnamefont {N.}~\bibnamefont {Zorbach}},
  \bibinfo {author} {\bibfnamefont {A.}~\bibnamefont {Koenigstein}}, \bibinfo
  {author} {\bibfnamefont {M.~J.}\ \bibnamefont {Steil}},\ and\ \bibinfo
  {author} {\bibfnamefont {S.}~\bibnamefont {Rechenberger}},\ }\href@noop {}
  {\bibinfo {title} {{Bosonic fluctuations in the $( 1 + 1 )$-dimensional
  Gross-Neveu(-Yukawa) model at varying $\mu$ and $T$ and finite $N$}}}
  (\bibinfo {year} {2021}),\ \Eprint {https://arxiv.org/abs/2108.10616}
  {arXiv:2108.10616 [hep-ph]} \BibitemShut {NoStop}%
\bibitem [{\citenamefont {Lenz}\ \emph {et~al.}(2022)\citenamefont {Lenz},
  \citenamefont {Mandl},\ and\ \citenamefont {Wipf}}]{Lenz:2021kzo}%
  \BibitemOpen
  \bibfield  {author} {\bibinfo {author} {\bibfnamefont {J.~J.}\ \bibnamefont
  {Lenz}}, \bibinfo {author} {\bibfnamefont {M.}~\bibnamefont {Mandl}},\ and\
  \bibinfo {author} {\bibfnamefont {A.}~\bibnamefont {Wipf}},\ }\bibfield
  {title} {\bibinfo {title} {{Inhomogeneities in the two-flavor chiral
  Gross-Neveu model}},\ }\href {https://doi.org/10.1103/PhysRevD.105.034512}
  {\bibfield  {journal} {\bibinfo  {journal} {Phys. Rev. D}\ }\textbf {\bibinfo
  {volume} {105}},\ \bibinfo {pages} {034512} (\bibinfo {year} {2022})},\
  \Eprint {https://arxiv.org/abs/2109.05525} {arXiv:2109.05525 [hep-lat]}
  \BibitemShut {NoStop}%
\bibitem [{\citenamefont {Lenz}\ and\ \citenamefont
  {Mandl}(2022)}]{Lenz:2021vdz}%
  \BibitemOpen
  \bibfield  {author} {\bibinfo {author} {\bibfnamefont {J.~J.}\ \bibnamefont
  {Lenz}}\ and\ \bibinfo {author} {\bibfnamefont {M.}~\bibnamefont {Mandl}},\
  }\bibfield  {title} {\bibinfo {title} {{Remnants of large-$N_\mathrm{f}$
  inhomogeneities in the 2-flavor chiral Gross-Neveu model}},\ }\href
  {https://doi.org/10.22323/1.396.0415} {\bibfield  {journal} {\bibinfo
  {journal} {PoS}\ }\textbf {\bibinfo {volume} {LATTICE2021}},\ \bibinfo
  {pages} {415} (\bibinfo {year} {2022})},\ \Eprint
  {https://arxiv.org/abs/2110.12757} {arXiv:2110.12757 [hep-lat]} \BibitemShut
  {NoStop}%
\bibitem [{\citenamefont {Horie}\ and\ \citenamefont
  {Nonaka}(2021)}]{Horie:2021wnn}%
  \BibitemOpen
  \bibfield  {author} {\bibinfo {author} {\bibfnamefont {K.}~\bibnamefont
  {Horie}}\ and\ \bibinfo {author} {\bibfnamefont {C.}~\bibnamefont {Nonaka}},\
  }\bibfield  {title} {\bibinfo {title} {{Inhomogeneous Phases in the Chiral
  Gross-Neveu Model on the Lattice}},\ }in\ \href@noop {} {\emph {\bibinfo
  {booktitle} {{38th International Symposium on Lattice Field Theory}}}}\
  (\bibinfo {year} {2021})\ \Eprint {https://arxiv.org/abs/2112.02261}
  {arXiv:2112.02261 [hep-lat]} \BibitemShut {NoStop}%
\bibitem [{\citenamefont {Ciccone}\ \emph {et~al.}(2022)\citenamefont
  {Ciccone}, \citenamefont {Di~Pietro},\ and\ \citenamefont
  {Serone}}]{Ciccone:2022zkg}%
  \BibitemOpen
  \bibfield  {author} {\bibinfo {author} {\bibfnamefont {R.}~\bibnamefont
  {Ciccone}}, \bibinfo {author} {\bibfnamefont {L.}~\bibnamefont {Di~Pietro}},\
  and\ \bibinfo {author} {\bibfnamefont {M.}~\bibnamefont {Serone}},\
  }\bibfield  {title} {\bibinfo {title} {{Inhomogeneous Phase of the Chiral
  Gross-Neveu Model}},\ }\href {https://doi.org/10.1103/PhysRevLett.129.071603}
  {\bibfield  {journal} {\bibinfo  {journal} {Phys. Rev. Lett.}\ }\textbf
  {\bibinfo {volume} {129}},\ \bibinfo {pages} {071603} (\bibinfo {year}
  {2022})},\ \Eprint {https://arxiv.org/abs/2203.07451} {arXiv:2203.07451
  [hep-th]} \BibitemShut {NoStop}%
\bibitem [{\citenamefont {Pisarski}\ \emph {et~al.}(2020)\citenamefont
  {Pisarski}, \citenamefont {Tsvelik},\ and\ \citenamefont
  {Valgushev}}]{Pisarski:2020dnx}%
  \BibitemOpen
  \bibfield  {author} {\bibinfo {author} {\bibfnamefont {R.~D.}\ \bibnamefont
  {Pisarski}}, \bibinfo {author} {\bibfnamefont {A.~M.}\ \bibnamefont
  {Tsvelik}},\ and\ \bibinfo {author} {\bibfnamefont {S.}~\bibnamefont
  {Valgushev}},\ }\bibfield  {title} {\bibinfo {title} {{How transverse thermal
  fluctuations disorder a condensate of chiral spirals into a quantum spin
  liquid}},\ }\href {https://doi.org/10.1103/PhysRevD.102.016015} {\bibfield
  {journal} {\bibinfo  {journal} {Phys. Rev. D}\ }\textbf {\bibinfo {volume}
  {102}},\ \bibinfo {pages} {016015} (\bibinfo {year} {2020})},\ \Eprint
  {https://arxiv.org/abs/2005.10259} {arXiv:2005.10259 [hep-ph]} \BibitemShut
  {NoStop}%
\bibitem [{\citenamefont {Nishimura}\ \emph {et~al.}(2015)\citenamefont
  {Nishimura}, \citenamefont {Ogilvie},\ and\ \citenamefont
  {Pangeni}}]{Nishimura:2014kla}%
  \BibitemOpen
  \bibfield  {author} {\bibinfo {author} {\bibfnamefont {H.}~\bibnamefont
  {Nishimura}}, \bibinfo {author} {\bibfnamefont {M.~C.}\ \bibnamefont
  {Ogilvie}},\ and\ \bibinfo {author} {\bibfnamefont {K.}~\bibnamefont
  {Pangeni}},\ }\bibfield  {title} {\bibinfo {title} {{Complex Saddle Points
  and Disorder Lines in QCD at finite temperature and density}},\ }\href
  {https://doi.org/10.1103/PhysRevD.91.054004} {\bibfield  {journal} {\bibinfo
  {journal} {Phys. Rev. D}\ }\textbf {\bibinfo {volume} {91}},\ \bibinfo
  {pages} {054004} (\bibinfo {year} {2015})},\ \Eprint
  {https://arxiv.org/abs/1411.4959} {arXiv:1411.4959 [hep-ph]} \BibitemShut
  {NoStop}%
\bibitem [{\citenamefont {Schindler}\ \emph {et~al.}(2020)\citenamefont
  {Schindler}, \citenamefont {Schindler}, \citenamefont {Medina},\ and\
  \citenamefont {Ogilvie}}]{Schindler:2019ugo}%
  \BibitemOpen
  \bibfield  {author} {\bibinfo {author} {\bibfnamefont {M.~A.}\ \bibnamefont
  {Schindler}}, \bibinfo {author} {\bibfnamefont {S.~T.}\ \bibnamefont
  {Schindler}}, \bibinfo {author} {\bibfnamefont {L.}~\bibnamefont {Medina}},\
  and\ \bibinfo {author} {\bibfnamefont {M.~C.}\ \bibnamefont {Ogilvie}},\
  }\bibfield  {title} {\bibinfo {title} {{Universality of Pattern Formation}},\
  }\href {https://doi.org/10.1103/PhysRevD.102.114510} {\bibfield  {journal}
  {\bibinfo  {journal} {Phys. Rev. D}\ }\textbf {\bibinfo {volume} {102}},\
  \bibinfo {pages} {114510} (\bibinfo {year} {2020})},\ \Eprint
  {https://arxiv.org/abs/1906.07288} {arXiv:1906.07288 [hep-lat]} \BibitemShut
  {NoStop}%
\bibitem [{\citenamefont {Schindler}\ \emph {et~al.}(2021)\citenamefont
  {Schindler}, \citenamefont {Schindler},\ and\ \citenamefont
  {Ogilvie}}]{Schindler:2021otf}%
  \BibitemOpen
  \bibfield  {author} {\bibinfo {author} {\bibfnamefont {M.~A.}\ \bibnamefont
  {Schindler}}, \bibinfo {author} {\bibfnamefont {S.~T.}\ \bibnamefont
  {Schindler}},\ and\ \bibinfo {author} {\bibfnamefont {M.~C.}\ \bibnamefont
  {Ogilvie}},\ }\bibfield  {title} {\bibinfo {title} {{$\mathcal PT$ symmetry,
  pattern formation, and finite-density QCD}},\ }\href
  {https://doi.org/10.1088/1742-6596/2038/1/012022} {\bibfield  {journal}
  {\bibinfo  {journal} {J. Phys. Conf. Ser.}\ }\textbf {\bibinfo {volume}
  {2038}},\ \bibinfo {pages} {012022} (\bibinfo {year} {2021})},\ \Eprint
  {https://arxiv.org/abs/2106.07092} {arXiv:2106.07092 [hep-lat]} \BibitemShut
  {NoStop}%
\bibitem [{\citenamefont {Schindler}\ \emph {et~al.}(2022)\citenamefont
  {Schindler}, \citenamefont {Schindler},\ and\ \citenamefont
  {Ogilvie}}]{Schindler:2021cke}%
  \BibitemOpen
  \bibfield  {author} {\bibinfo {author} {\bibfnamefont {M.~A.}\ \bibnamefont
  {Schindler}}, \bibinfo {author} {\bibfnamefont {S.~T.}\ \bibnamefont
  {Schindler}},\ and\ \bibinfo {author} {\bibfnamefont {M.~C.}\ \bibnamefont
  {Ogilvie}},\ }\bibfield  {title} {\bibinfo {title} {{Finite-density QCD,
  $\mathcal{PT}$ symmetry, and exotic phases}},\ }\href
  {https://doi.org/10.22323/1.396.0555} {\bibfield  {journal} {\bibinfo
  {journal} {PoS}\ }\textbf {\bibinfo {volume} {LATTICE2021}},\ \bibinfo
  {pages} {555} (\bibinfo {year} {2022})},\ \Eprint
  {https://arxiv.org/abs/2110.07761} {arXiv:2110.07761 [hep-lat]} \BibitemShut
  {NoStop}%
\bibitem [{\citenamefont {Winstel}\ and\ \citenamefont
  {Valgushev}(2024)}]{Winstel:2024qle}%
  \BibitemOpen
  \bibfield  {author} {\bibinfo {author} {\bibfnamefont {M.}~\bibnamefont
  {Winstel}}\ and\ \bibinfo {author} {\bibfnamefont {S.}~\bibnamefont
  {Valgushev}},\ }\bibfield  {title} {\bibinfo {title} {{Lattice study of
  disordering of inhomogeneous condensates and the Quantum Pion Liquid in
  effective $O(N)$ model}},\ }in\ \href@noop {} {\emph {\bibinfo {booktitle}
  {{Excited QCD 2024 Workshop}}}}\ (\bibinfo {year} {2024})\ \Eprint
  {https://arxiv.org/abs/2403.18640} {arXiv:2403.18640 [hep-lat]} \BibitemShut
  {NoStop}%
\bibitem [{\citenamefont {Haensch}\ \emph {et~al.}(2023)\citenamefont
  {Haensch}, \citenamefont {Rennecke},\ and\ \citenamefont {von
  Smekal}}]{Haensch:2023sig}%
  \BibitemOpen
  \bibfield  {author} {\bibinfo {author} {\bibfnamefont {M.}~\bibnamefont
  {Haensch}}, \bibinfo {author} {\bibfnamefont {F.}~\bibnamefont {Rennecke}},\
  and\ \bibinfo {author} {\bibfnamefont {L.}~\bibnamefont {von Smekal}},\
  }\href@noop {} {\bibinfo {title} {{Medium Induced Mixing and Critical Modes
  in QCD}}} (\bibinfo {year} {2023}),\ \Eprint
  {https://arxiv.org/abs/2308.16244} {arXiv:2308.16244 [hep-ph]} \BibitemShut
  {NoStop}%
\bibitem [{\citenamefont {Basar}\ and\ \citenamefont
  {Dunne}(2008{\natexlab{a}})}]{Basar:2008im}%
  \BibitemOpen
  \bibfield  {author} {\bibinfo {author} {\bibfnamefont {G.}~\bibnamefont
  {Basar}}\ and\ \bibinfo {author} {\bibfnamefont {G.~V.}\ \bibnamefont
  {Dunne}},\ }\bibfield  {title} {\bibinfo {title} {{Self-consistent
  crystalline condensate in chiral Gross-Neveu and Bogoliubov-de Gennes
  systems}},\ }\href {https://doi.org/10.1103/PhysRevLett.100.200404}
  {\bibfield  {journal} {\bibinfo  {journal} {Phys. Rev. Lett.}\ }\textbf
  {\bibinfo {volume} {100}},\ \bibinfo {pages} {200404} (\bibinfo {year}
  {2008}{\natexlab{a}})},\ \Eprint {https://arxiv.org/abs/0803.1501}
  {arXiv:0803.1501 [hep-th]} \BibitemShut {NoStop}%
\bibitem [{\citenamefont {Basar}\ and\ \citenamefont
  {Dunne}(2008{\natexlab{b}})}]{Basar:2008ki}%
  \BibitemOpen
  \bibfield  {author} {\bibinfo {author} {\bibfnamefont {G.}~\bibnamefont
  {Basar}}\ and\ \bibinfo {author} {\bibfnamefont {G.~V.}\ \bibnamefont
  {Dunne}},\ }\bibfield  {title} {\bibinfo {title} {{A Twisted Kink Crystal in
  the Chiral Gross-Neveu model}},\ }\href
  {https://doi.org/10.1103/PhysRevD.78.065022} {\bibfield  {journal} {\bibinfo
  {journal} {Phys. Rev. D}\ }\textbf {\bibinfo {volume} {78}},\ \bibinfo
  {pages} {065022} (\bibinfo {year} {2008}{\natexlab{b}})},\ \Eprint
  {https://arxiv.org/abs/0806.2659} {arXiv:0806.2659 [hep-th]} \BibitemShut
  {NoStop}%
\bibitem [{\citenamefont {Dunne}\ and\ \citenamefont
  {Thies}(2013)}]{Dunne:2013xta}%
  \BibitemOpen
  \bibfield  {author} {\bibinfo {author} {\bibfnamefont {G.~V.}\ \bibnamefont
  {Dunne}}\ and\ \bibinfo {author} {\bibfnamefont {M.}~\bibnamefont {Thies}},\
  }\bibfield  {title} {\bibinfo {title} {{Time-Dependent Hartree-Fock Solution
  of Gross-Neveu models: Twisted Kink Constituents of Baryons and Breathers}},\
  }\href {https://doi.org/10.1103/PhysRevLett.111.121602} {\bibfield  {journal}
  {\bibinfo  {journal} {Phys. Rev. Lett.}\ }\textbf {\bibinfo {volume} {111}},\
  \bibinfo {pages} {121602} (\bibinfo {year} {2013})},\ \Eprint
  {https://arxiv.org/abs/1306.4007} {arXiv:1306.4007 [hep-th]} \BibitemShut
  {NoStop}%
\bibitem [{\citenamefont {Koenigstein}(2023)}]{Koenigstein:2023wso}%
  \BibitemOpen
  \bibfield  {author} {\bibinfo {author} {\bibfnamefont {A.}~\bibnamefont
  {Koenigstein}},\ }\emph {\bibinfo {title} {Non-perturbative aspects of
  (low-dimensional) quantum field theories}},\ \href
  {https://doi.org/10.21248/gups.74658} {\bibinfo {type} {Phd thesis}},\
  \bibinfo  {school} {Universit{\"a}tsbibliothek Johann Christian Senckenberg}
  (\bibinfo {year} {2023})\BibitemShut {NoStop}%
\bibitem [{\citenamefont {Le~Bellac}(1991)}]{LeBellac:1991cq}%
  \BibitemOpen
  \bibfield  {author} {\bibinfo {author} {\bibfnamefont {M.}~\bibnamefont
  {Le~Bellac}},\ }\href@noop {} {\emph {\bibinfo {title} {{Quantum and
  statistical field theory}}}}\ (\bibinfo  {publisher} {Oxford University
  Press},\ \bibinfo {address} {Oxford, England, UK},\ \bibinfo {year} {1991})\
  \bibinfo {note} {[Translation by G.~Barton]}\BibitemShut {NoStop}%
\bibitem [{\citenamefont {Kapusta}\ and\ \citenamefont
  {Gale}(2011)}]{Kapusta:2006pm}%
  \BibitemOpen
  \bibfield  {author} {\bibinfo {author} {\bibfnamefont {J.~I.}\ \bibnamefont
  {Kapusta}}\ and\ \bibinfo {author} {\bibfnamefont {C.}~\bibnamefont {Gale}},\
  }\href {https://doi.org/10.1017/CBO9780511535130} {\emph {\bibinfo {title}
  {{Finite-temperature field theory: Principles and applications}}}},\
  Cambridge Monographs on Mathematical Physics\ (\bibinfo  {publisher}
  {Cambridge University Press},\ \bibinfo {year} {2011})\BibitemShut {NoStop}%
\bibitem [{\citenamefont {Thies}(2022{\natexlab{b}})}]{Thies:2022kuv}%
  \BibitemOpen
  \bibfield  {author} {\bibinfo {author} {\bibfnamefont {M.}~\bibnamefont
  {Thies}},\ }\bibfield  {title} {\bibinfo {title} {{Tricritical curve of
  massive chiral Gross-Neveu model with isospin}},\ }\href
  {https://doi.org/10.1103/PhysRevD.106.056026} {\bibfield  {journal} {\bibinfo
   {journal} {Phys. Rev. D}\ }\textbf {\bibinfo {volume} {106}},\ \bibinfo
  {pages} {056026} (\bibinfo {year} {2022}{\natexlab{b}})},\ \Eprint
  {https://arxiv.org/abs/2207.14503} {arXiv:2207.14503 [hep-th]} \BibitemShut
  {NoStop}%
\bibitem [{\citenamefont {{Wolfram Research{,}
  Inc.}}(2023)}]{Mathematica:13.0}%
  \BibitemOpen
  \bibfield  {author} {\bibinfo {author} {\bibnamefont {{Wolfram Research{,}
  Inc.}}},\ }\href {https://www.wolfram.com/mathematica} {\bibinfo {title}
  {{Mathematica, {V}ersion 13.0}}} (\bibinfo {year} {2023})\BibitemShut
  {NoStop}%
\bibitem [{\citenamefont {Horvát}(2020)}]{Horvat:matex}%
  \BibitemOpen
  \bibfield  {author} {\bibinfo {author} {\bibfnamefont {S.}~\bibnamefont
  {Horvát}},\ }\href
  {http://szhorvat.net/pelican/latex-typesetting-in-mathematica.html} {\bibinfo
  {title} {{MaTeX}}} (\bibinfo {year} {2020}),\ \bibinfo {note} {[Online;
  accessed 2020.10.08]}\BibitemShut {NoStop}%
\bibitem [{\citenamefont {GitHub}\ and\ \citenamefont
  {LP}(2021)}]{GitHubCopilot2021}%
  \BibitemOpen
  \bibfield  {author} {\bibinfo {author} {\bibfnamefont {I.}~\bibnamefont
  {GitHub}}\ and\ \bibinfo {author} {\bibfnamefont {O.}~\bibnamefont {LP}},\
  }\href@noop {} {\bibinfo {title} {Github copilot}},\ \bibinfo {howpublished}
  {\url{https://copilot.github.com}} (\bibinfo {year} {2021}),\ \bibinfo {note}
  {accessed: 2024-04-18}\BibitemShut {NoStop}%
\end{thebibliography}%


\end{document}